%Paper: astro-ph/9506021
%From: mail <FNUSOV@WEIZMANN.WEIZMANN.AC.IL>
%Date: Sun, 04 Jun 95 09:55:29 +0300

\magnification=1200

%\nopagenumbers
\hsize=165truemm
\vsize=235truemm
\voffset=2\baselineskip
\parskip=4truept plus1.3truept minus.7truept

\def\section#1\par{\bigskip\bigskip\goodbreak\centerline
{\bf#1}\nobreak\medskip\nobreak\par\noindent}

\def\subsection#1\par{\bigskip\goodbreak\noindent{\it#1}
\nobreak\smallskip\nobreak\par\noindent}

\def\ga{\lower.4ex\hbox{$\;\buildrel >\over{\scriptstyle\sim}\;$}}
\def\la{\lower.4ex\hbox{$\;\buildrel <\over{\scriptstyle\sim}\;$}}
\def\hang{\par\noindent\hangindent\parindent}
\def\ctrline{\centerline}

\def\illustration #1 by #2 (#3){\hfill{
 \vbox to #2{
  \hrule width #1 height 0pt depth 0pt
   \vfill
    \special{illustration #3}}}\hfill}

\def\scaledillustration #1 by #2 (#3 scaled #4){{
     \dimen0=#1 \dimen1=#2
      \divide\dimen0 by 1000 \multiply\dimen0 by #4
      \divide\dimen1 by 1000 \multiply\dimen1 by #4
       \illustration \dimen0 by \dimen1 (#3 scaled #4)}}

\def\et{{\it et al}.\ }

\def\and{and\ }%{\&\ }

\def\tinycaps#1{{\rm\scriptscriptstyle#1}}
\def\S{{\tinycaps S}}
\def\RS{\tinycaps {RS}}
\def\GJ{\tinycaps {GJ}}
\def\A{\tinycaps A}
\def\B{\tinycaps B}
\def\F{\tinycaps F}
\def\D{\tinycaps D}
\def\E{\tinycaps E}

\vskip-12pt
\hfill
%Version of 3 March 1995
\medskip

\ctrline{\bf PULSARS WITH STRONG MAGNETIC FIELDS: POLAR GAPS,}
\medskip
\ctrline{\bf BOUND PAIR CREATION AND NONTHERMAL LUMINOSITIES}

\bigskip
\ctrline{V.V. Usov}
\ctrline{Physics Department, Weizmann Institute}
\ctrline{Rehovot 76100, ISRAEL}
\medskip
\ctrline{and D.B. Melrose}
\ctrline{Research Centre for Theoretical Astrophysics}
\ctrline{School of Physics, University of
Sydney, NSW 2006, AUSTRALIA}

\bigskip%\baselineskip 20pt
\ctrline{\bf Abstract}

Modifications to polar-gap models for pulsars are discussed for the
case where the surface magnetic field, $B_\S$, of the neutron star
is strong. For $B\ga4\times10^8\rm\,T$, the curvature $\gamma$-quanta
emitted tangentially to the curved force lines of the magnetic field are
captured near the threshold of bound pair creation and are channelled
along the magnetic field as bound electron-positron pairs (positronium).
The stability of such bound pairs against ionization by the parallel
electric field, $E_\parallel$, in the polar cap, and against
photoionization is discussed. Unlike free pairs, bound pairs do not screen
$E_\parallel$ near the neutron star. As a consequence, the energy flux in
highly relativistic particles and high-frequency (X-ray and/or
$\gamma$-ray) radiation from the polar gaps can be much greater than in
the absence of positronium formation. We discuss this enhancement for (a)
Arons-type models, in which particles flow freely from the surface, and
find any enhancement to be modest, and (b) Ruderman-Sutherland-type
models, in which particles are tightly bound to the surface, and find that
the enhancement can be substantial. In the latter case we argue for a
self-consistent, time-independent model in which partial screening of
$E_\parallel$ maintains it close to the threshold value for field
ionization of the bound pairs, and in which a reverse flux of accelerated
particles maintains the polar cap at a temperature such that thermionic
emission supplies the particles needed for this screening. This model
applies only in a restricted range of periods, $P_2<P<P_1$, and it
implies an energy flux in high-energy particles that can correspond to a
substantial fraction of the spin-down power of the pulsar.

Nonthermal, high-frequency radiation has been observed from six radio
pulsars and Geminga is usually included as a seventh case. The nonthermal
luminosity can be higher than can be explained in terms of conventional
polar-gap and outer-gap models. The self-consistent polar-gap model
proposed here alleviates this difficulty, provided the magnetic field
satisfies $B\ga4\times10^8\rm\,T$ (which is so for five of these pulsars,
and plausibly for the other two if a modest nondipolar component is
assumed), and the surface temperature (in the absence of heating by the
reverse flux) satisfies $T_\S\la 10^6\rm\,K$, so that thermionic
emission from the surface is unimportant. It is argued that sufficient
power is available to explain the observed high-frequency radiation of
most of these pulsars. However, the Crab and PSR 0540--69 have periods
$P<P_2$, and we suggest that an outer-gap model is more appropriate for
these.

\vfill\eject

\section{1. INTRODUCTION}

The creation of electron-positron pairs by decay of $\gamma$ rays
as they propagate across magnetic field lines is an essential
ingredient in the population of a pulsar magnetosphere with plasma.
In the absence of plasma, a strong electric fields ${\bf E}$
results from the rotation of the magnetized neutron star (Goldreich
\and Julian 1969). The vacuum field has a nonzero parallel component,
$E_\parallel=({\bf E\cdot B})/\vert{\bf B}\vert$ along the magnetic field
${\bf B}$, and this $E_\parallel$ can accelerate {\it primary\/} particles
to ultrarelativistic energies (e.g., the review by Michel 1991). The
source of the primary particles is different in different models. In the
absence of free ejection from the stellar surface, the primary particles
come from a pair cascade initiated, for example, by decay of stray
$\gamma$ rays into pairs (Ruderman \and Sutherland 1975). In such models
the vacuum $E_\parallel$ is present immediately above the surface and is
said to form a vacuum gap. In models with free ejection of particles from
the stellar surface, these particles screen $E_\parallel$
immediately above the surface, and set up the corotation electric field
${\bf E}_{\rm rot}=-({\bf\Omega}\times{\bf r})\times{\bf B}$, where
$\bf\Omega$ is the angular velocity of rotation. The divergence of
${\bf E}_{\rm rot}$ requires a charge density, $en_\GJ$, where $n_\GJ$ is
the Goldreich-Julian density (Goldreich \and Julian 1969). In such models
an $E_\parallel$ develops, forming a vacuum-gap-like region above the
surface, due to the actual charge density deviating increasingly with
increasing height from the Goldreich-Julian value, $n\ne n_\GJ$. In either
model, the primary particles emit $\gamma$ rays, due to curvature emission
and other processes. Some of these $\gamma$ rays are absorbed in the
magnetic field by creating free electron-positron pairs
$(\gamma+B\rightarrow e^++e^-+B)$. Provided these {\it secondary\/}
particles are created in the gap, the $E_\parallel$ field can accelerate
the electron and the positron in opposite directions, allowing a net
charge density to build up, and it is this charge density of the secondary
particles that screens $E_\parallel$. Regions where $E_\parallel$ is
unscreened are called gaps, and a gap that forms near the magnetic poles
of the pulsar is called a {\it polar gap}. The height, $H$, of a polar gap
is defined by the height above the stellar surface where the charge
density due to the charge-separated pairs is sufficient to screen
$E_\parallel$. Besides polar-gap models there are also outer-gap models
(e.g., Cheng, Ho \and Ruderman 1986a,b), where the region with
$E_\parallel\ne0$ occurs far from the stellar surface, where the magnetic
field is much weaker than in the polar-cap regions.
In comparing the slot-gap with polar-gap and outer-gap models, we note
that the slot gaps form on the boundary of the closed $B$-field
region (Arons 1983). Primary particles which are ejected from the neutron
star surface and are accelerated in the slot gap gain the main part of
their energy far from the neutron star. In this respect, slot gaps
differ significantly from the polar gaps and may be regarded as
intermediate between polar gaps and outer gaps.

Polar gaps are thought to be a source of the energy for nonthermal
radiation of pulsars (Sturrock 1971; Ruderman \and Sutherland 1975;
Michel 1975; Arons 1979; Arons \and Scharlemann 1979; Cheng \and
Ruderman 1980; Arons 1981; Mestel 1981, 1993; Mestel \et 1985;
Fitzpatrick \and Mestel 1988a,b; Shibata 1991). In particular, in
most models the radio emission is attributed to processes in polar
gaps. However, the radio luminosities of the pulsars are small
compared with the total power loss estimated from the
observed spin down and associated rotational energy loss ($\la10^{-5}$).
Most of the power loss is attributed to relativistic particles created as
secondary pairs in the polar gaps and escaping in a relativistic wind
(Ruderman \and Sutherland 1975; Arons 1979; Michel 1991). In most
pulsars the only nonthermal radiation observed is in the radio range
and, apart from the slowing down, the large inferred power loss has
no direct observational signature. However, there are a few radio
pulsars that do have high-frequency emission (in X rays and/or
$\gamma$ rays, we do not distinguish between them in this paper) with a
nonthermal luminosity that is a substantial fraction of the inferred
rotational power loss. Nonthermal high-frequency radiation requires high
energy particles, and it is widely assumed that the high-frequency
luminosity correlates with the power in primary particles (e.g., Harding
\and Daugherty 1993). The power in the primary particles, ${\dot N}_{\rm
prim}e\Delta\varphi$, is limited by the rate of injection of such
particles, ${\dot N}_{\rm prim}$, which is less than $n_\GJ c$ times the
area of the polar cap, and the potential energy, $e\Delta\varphi\sim
eE_\parallel H$, that particles gain in crossing the polar gap. The
maximum potential difference across the polar gap is when there is no
screening at all: $$\Delta\varphi_{\rm max}=
{\Omega^2B^{\rm d}_\S R^3\over 2c^2}\simeq
(6.6\times10^{14}{\rm\,V})
\left(B^{\rm d}_\S\over10^8\rm\,T\right)
\left({P\over0.1{\rm\,s}}\right)^{-2}\,,
\eqno(1.1)$$

\noindent where $B^{\rm d}_\S$ is the dipole component of the magnetic
field at the magnetic pole on the neutron star surface, $\Omega=2\pi/P$
is the angular speed of rotation of the neutron star and $R\simeq10^4\rm
\,m$ is the radius of the neutron star. For $\Delta\varphi=\Delta
\varphi_{\rm max}$ the implied maximum value for power in primary
particle is equal (within a factor of order unity) to the rotational
energy loss. For existing polar-gap models the value of $\Delta\varphi$
is, as a rule, considerably less than $\Delta\varphi_{\rm max}$.
In Arons-type models, in which particles flow freely from the surface,
this is due both (a)~to $E_\parallel$ being much less than the vacuum
value due to screening by the primary particles that flow freely from the
surface, and (b)~to the height being less than the maximum value,
$H\sim R$, for such a model due to screening by secondary pairs. In
Ruderman-Sutherland-type models, in which particles are tightly bound to
the surface, $\Delta\varphi\ll\Delta\varphi_{\rm max}$ results from
$H\ll\Delta R_p$ due to screening by secondary pairs, where $\Delta R_p$
is the radius of the polar cap which is the maximum value of $H$ in
this kind of polar-gap model.

Strong high-frequency radiation is observed from the galactic pulsars PSR
0531+21, PSR 0833--45, PSR 1055--52, PSR 1509--58 and PSR 1706--44, and
PSR 0540--69 in the Large Magellanic Cloud (e.g., the review by Ulmer
1994). The observed radiated power for all these pulsars is concentrated
in the X-ray or $\gamma$-ray ranges. The $\gamma$-ray pulsar Geminga is
probably also a radio pulsar (Halpern \and Holt 1992) which is ``radio
quiet'' because its radio beam does not intersect the Earth (Ozernoy \and
Usov 1977). The high $\gamma$-ray luminosities are inferred assuming
emission into a large solid angle, $\sim2\pi\rm\,sterad$ (Ulmer 1994). For
some of the the pulsars, notably for PSR 1055--52, among the present
polar-gap models the only one that appears capable of explaining the
inferred very high luminosities is that of Sturrock (1971). In the
Sturrock model the potential drop has the maximum possible value, and the
nonthermal power is of order the rotational power loss. However, the
Sturrock model is not self-consistent because it neglects the finite limit
on the height of the polar gap due to screening. In the Sturrock model the
free electron-positron pairs created by the curvature $\gamma$-quanta
inside the polar gap are assumed not to feel the electric field
$E_\parallel$ and not to screen this field so that the full potential drop
$\Delta\varphi_{\rm max}$ is implicitly available. This internal
inconsistency implies that the Sturrock model should not be used for
detailed estimates of the pulsar luminosities.

Ruderman \and Sutherland (1975) were the first to develop a
self-consistent polar-gap model in which the screening of the
electric field $E_\parallel$ by the electron-positron pairs created
in it is taken into account. Consideration of this screening led
Ruderman \and Sutherland (1975) to conclude that the potential across
the polar gap cannot exceed $\Delta\varphi_\RS\simeq$ a few
$\times10^{12}\rm\,V$. This upper limit on the potential across the
polar gap is valid for any polar-gap model in which free pairs are
created by $\gamma$-quanta absorption in the magnetic field. For young,
rapidly-rotating pulsars with typical parameters, $P\la0.1\rm\,s$
and $B_\S^{\rm d}\ga10^8\rm\,T$, one has
$\Delta\varphi_\RS\la10^{-3}\,\Delta\varphi_{\rm max}$. The total power
in primary particles is proportional to $\Delta \varphi$, and the
maximum luminosity expected in all existing polar-gap models (Ruderman
\and Sutherland 1975; Arons 1979, 1981; Arons \and Scharlemann 1979; Cheng
\and Ruderman 1980; Mestel \et 1985) is smaller than the inferred
nonthermal luminosity in X-rays and $\gamma$ rays (section 5).

In all existing polar-gap models, the formation of pairs in the pulsar
magnetospheres is due to the single-photon mechanism, $\gamma+B\rightarrow
e^++e^-+B$, and the resulting pairs are free. However, this assumption
that the pairs are free is not valid if the magnetic field is strong
enough, specifically for $B>0.1\,B_{\rm cr}$, where
$B_{\rm cr}=m^2c^2/e\hbar=4.4\times10^9{\rm\,T}$ is known as the critical
field. In such a strong magnetic field, the curvature $\gamma$-quanta
emitted tangentially to the curved force lines of the magnetic field are
captured near the threshold of bound pair creation and are then channelled
along the magnetic field as positronium, that is, as bound pairs (Shabad
\and Usov 1982, 1985, 1986; Herold, Ruder \and Wunner 1985; Usov \and
Shabad 1985; M\'esz\'aros 1992; Shabad 1992 and references therein). This
positronium may be stable in the polar gaps against both the
ionizing action of the electric field and against photo-ionization
(Shabad \and Usov 1985; Bhatia, Chopra \and Panchapakesan 1988, 1992).
Unlike free pairs, such bound pairs do not screen the electric field
$E_\parallel$ near the pulsar. Screening requires a net charge density,
which can build up due to free pairs being separated by $E_\parallel$, but
cannot build up if the pairs remain bound. As a result the height of the
gap, which is determined by the height at which screening becomes
important, is greater than it would be in the absence of formation of
positronium. The assumption that the power in primary particles is
proportional to $H$ in polar-gap models, implies that this total
luminosity increases when positronium formation becomes important (Usov
\and Shabad 1985). We argue in section 5 that the nonthermal
luminosity may even become comparable with the maximum possible, as in the
model of Sturrock (1971).

Our primary objective in this paper is to discuss how models of pulsar
electrodynamics need to be modified when the magnetic field is strong
enough for creation of bound pairs to dominate over the formation of free
pairs. First, however, we review several important preliminary aspects of
the problem. The rate of injection of the primary particles depends on the
properties of the matter at the surface of a magnetic neutron star.
The particle outflow from the neutron star surface is discussed in
section~2. Particle acceleration and generation of curvature
$\gamma$ rays in the pulsar magnetospheres are described in
section~3. Propagation of $\gamma$ rays, including photon splitting,
and creation of electron-positron pairs in a strong magnetic field
are discussed in section~4. In section~5, the polar-gap model is
developed for pulsars with strong magnetic fields at their surface,
$B_\S > 0.1\,B_{\rm cr}$, and the maximum value of the nonthermal
luminosity is estimated for such a pulsar. The interpretation of the
observational data and some theoretical predictions on the
nonthermal radiation of pulsars with $B_\S > 0.1\,B_{\rm cr}$ are
also given in section~5. A discussion and summary are presented in
section~6.

\section{2. THE NEUTRON STAR SURFACE AND PARTICLE EJECTION}

The electric field distribution and particle acceleration in
the polar gaps of pulsars depend on the character of particle
outflow from the pulsar surface. In some models the binding to
the surface is assumed sufficiently strong that no particles are ejected
from the surface, and in other models the binding is assumed sufficiently
weak that particles (either electrons or ions) flow freely from the
stellar surface. Particle ejection from the surface depends on the
composition of the neutron star surface, the structure of the surface with
a strong magnetic field and the surface temperature $T_\S$. In this
section, after a brief review of polar-gap models we discuss the
properties stellar surface and their effect on the binding or ejection of
ions. We then discuss models involving free ejection of electrons and free
ejection of ions.

\subsection{2.1. Polar-gap models}

There are several kinds of polar-gap model. These may be classified in two
ways: whether ions or electrons tend to be ejected from the surface, and
whether $E_\parallel$ is zero or nonzero at the stellar surface.

The sign of the charge of the particles that tend to be ejected from the
neutron star surface depends on the sign of ${\bf \Omega \cdot B}$.
Electrons tend to be ejected for ${\bf\Omega\cdot B}>0$ and ions for
${\bf\Omega\cdot B}<0$. The argument for this (Goldreich \and Julian 1969;
Ruderman \and Sutherland 1975; Arons 1979, 1981) is that plasma tends to
corotate with the star, and the divergence of the corotation electric
field implies a charge density, $en_\GJ$, where
$$n_\GJ={\vert {\bf \Omega
\cdot B}\vert \over 2\pi ce}\simeq
(7\times10^{17}{\rm\,m}^{-3})\left({B\over 10^8\rm\,T}\right)\left(
{P\over 0.1\,{\rm s}}\right)^{-1}\cos \Theta \,,
\eqno(2.1)$$

\noindent is the Goldreich-Julian density, and
$\Theta$ is the angle between the angular velocity ${\bf \Omega}$
and the magnetic field ${\bf B}$. Thus the tendency to corotate requires
positive charges above the polar caps for ${\bf\Omega\cdot B}<0$ and
negative charges over the polar caps for ${\bf\Omega\cdot B}>0$. The
required charge density may be produced in two ways: ejection of particles
with the appropriate sign of the charge from the stellar surface, or
acceleration (by $E_\parallel$) from the magnetosphere back to the stellar
surface of charges of the opposite sign.

The most familiar model in which there is no ejection of particles from
the stellar surface is that of Ruderman \and Sutherland (1975). This
model applies only to neutron stars with ${\bf\Omega}\cdot{\bf B}<0$
(neutron stars with ${\bf\Omega}\cdot{\bf B}>0$ are assumed not to be
pulsars). The absence of ion flow from the surface is attributed to an
assumed strong binding of iron nuclei. In the Ruderman-Sutherland model,
the field $E_\parallel$ is maximum at the surface and decreases with
distance. All the particles in the pulsar magnetosphere are created from
pair production through $\gamma$-ray absorption. The magnetosphere
is populated by a pair plasma created near the polar cap, and
the (positive) Goldreich-Julian charge density is created by electrons
being accelerated back to the stellar surface, leaving a net excess (and
outward flux) of positrons. A model with somewhat similar properties is
that of Beskin, Gurevich \and Istomin (1986) in which the pair plasma is
assumed to be created in a postulated double layer immediately above the
surface on the star.

In contrast, in the model of Arons (1979, 1981) it is assumed that
charged particles (electrons or ions depending on the sign of
${\bf\Omega\cdot B}$) flow freely from the neutron star surface. In this
model the electric field $E_\parallel$ is equal to zero at the surface,
and increases with distance above the surface. In the absence of any
additional source of charge, conservation of the current implies a current
density $J\propto B$. For so-called {\it favorably\/} curved field lines,
the ratio of the charge density ($n=J/ec\propto B$) to the
Goldreich-Julian charge density ($n_\GJ\propto{\bf\Omega}\cdot{\bf B}/
\Omega )$ decreases with distance above the stellar surface. This implies
that the screening of $E_\parallel$ is incomplete, and $E_\parallel$
increases toward the vacuum value. The resulting $E_\parallel$, which is
smaller than that in the Ruderman-Sutherland model, causes the ejected
electrons or ions to become highly relativistic and causes a net flux of
pair-produced electrons or positrons back to the stellar surface to allow
the charge density to be maintained at the Goldreich-Julian value. Mestel
\et (1985) pointed out that for the unfavorably curved field lines, along
which $n/n_\GJ$ increases, an $E_\parallel$ of opposite sign to the vacuum
field develops and plays essentially the same role as the $E_\parallel$ in
the Arons model.  A third kind of polar-gap model (Cheng \and Ruderman
1980) is an intermediate case where the particles flow from the pulsar
surface but not freely. In such a model the field $E_\parallel$ is nonzero
at the pulsar surface but it is smaller than in the model of Ruderman \and
Sutherland (1975).

In the discussion below, we use the Ruderman-Sutherland and Arons models
for illustrative purposes.

\subsection{2.2. Composition of the neutron star surface}

The strength of the binding of ions to the stellar surface is an important
ingredient in pulsar models with ${\bf\Omega\cdot B}<0$, and this binding
depends on the ionic composition. We note the following arguments for and
against the composition being primarily Fe$^{56}$.

\noindent
1). Rosen \and Cameron (1972) showed that the outermost layers of the
neutron star atmosphere, in which no mass is being ejected, are composed
almost entirely of He$^4$ with trace amounts of Fe$^{56}$. The He$^4$ is
produced by photodissociation of the iron-peak nuclei (assumed to be
the major initial constituent) at the high temperatures encountered
during collapse to the neutron star state. The total mass of the He$^4$
at the neutron star surface is
$\Delta M_{\rm He}\simeq1.3\times10^{18}\,{\rm kg}\sim 10^{-12}M_\odot$.
However, the luminosity of a very young neutron star is highly
super-Eddington, and so radiation pressure drives matter away from the
neutron star surface. After the loss of about $10^{-12}M_\odot$, only
Fe$^{56}$ remains at the surface (Rosen \and Cameron 1972). For young
neutron stars the mass-loss rate is as high as
$\sim0.005\,M_\odot\rm\,s^{-1}$, and the mass ejected from the neutron
star surface is of order $10^{-3}$ -- $10^{-2}M_\odot$ (Woosley \and Baron
1992; Levinson \and Eichler 1993). Thus the ejected mass is
$\gg\Delta M_{\rm He}$, suggesting that the neutron
star surface consists almost entirely of Fe$^{56}$.

\noindent
2). The composition of the polar caps of pulsars may differ substantially
from the composition of the main part of the neutron star surface due to
the bombardment by energetic particles from the magnetosphere. Pair
particles created near the top edge of the polar gap and accelerated by
the $E_\parallel$ field back to the stellar surface form showers as they
cross the surface (Cheng \and Ruderman 1977; Jones 1978, 1979). Protons and
spallation nuclei are produced at the polar caps of pulsars by hadronic
photoabsorption of shower photons. The flux of reversed electrons (or
positrons) through unit surface area of the polar cap is
$$F_r = \xi n_\GJ c\,,
\eqno(2.2)$$

\noindent where $\xi$ is a dimensionless parameter,
with $\xi\sim1$ for the model of Ruderman \and Sutherland (1975)
and $\xi\sim(\Omega R/c)\simeq2\times10^{-3}(P/0.1\rm\,s)^{-1}\ll1$
for the model of Arons (1979, 1981). The energy of reversed particles
is $\Gamma_rmc^2$ with $\Gamma_r\sim10^7$. Shower $\gamma$ rays penetrate
into the neutron star surface layers to $\sim 30$ radiation lengths (Jones
1978; Bogovalov \and Kotov 1989) and break up atomic nuclei. In the case of
Fe$^{56}$, for which the radiation length is
$l_r\simeq140\rm\,kg\,m^{-2}$, 30 radiation lengths corresponds to a
column density $\sigma=30l_r\simeq4.2\times10^3\rm\,kg\,m^{-2}$. The mean
number of photoabsorption events per one reversed particle with energy
$\sim10^7mc^2$ is $\sim10^3$ (Hayward 1965; Jones 1978). Hence, from
equations (2.1) and (2.2), the characteristic time for spallation of all
the Fe$^{56}$ in the surface layer, $\tau=(\sigma/{\rm A}m_p)/10^3F_r$
with ${\rm A}=56$, reduces to $$\tau <
(2{\rm\,s})\,\xi^{-1} \left({B_\S\over
10^8{\rm\,T}}\right)^{-1}
\left({P\over 1\,{\rm s}}\right)
(\cos\Theta)^{-1}\,,
\eqno(2.3)$$

\noindent which is very much shorter than the age of any pulsar. This
suggests that all the Fe$^{56}$ should be destroyed in the
surface layers of the polar caps.

\noindent
3). There are several contrary arguments to the foregoing.  (a)~The most
important photoabsorption reaction is the formation of the giant dipole
resonance (e.g., Hayward 1965; Jones 1978), which, for iron-group nuclei,
decays predominantly through neutron emission. Most of the photons with
energies of $\sim 15$ -- $30\rm\,MeV$ mainly responsible for spallation of
Fe$^{56}$ are generated in the deep layer, $\ga10\,l_r$,  (e.g., Jones
1978; Bogovalov \and Kotov 1989). This suggests that the surface layer
itself is relatively unaffected by the bombardment. (b)~The resulting
neutrons are distributed more or less uniformly inside the surface layer
due to diffusion (Jones 1978), and are recaptured by the other nuclei.
Neutron capture can reform Fe$^{56}$, so that it is not permanently
destroyed.  (c)~If the magnetic field is strong enough,
$B_\S>5\times10^{8}$ T, so that solid magnetic metal is formed at the
pulsar surface (see below), both the generation of photons with energies of
$\sim15$ -- $30\rm\,MeV$ and  spallation of Fe$^{56}$ may be
suppressed over a few $l_r$ due to the Landau-Pomeranchuk effect (Rozental
\and Usov 1985), that is, due to the suppression of the formation of softer
photons and pairs in the strong magnetic field. (d)~For pulsars with
${\bf\Omega\cdot B}<0$ at the polar caps, so that ions are the particles
ejected from the stellar surface, ejection of ions can dominate over
spallation. If so, the spallation products are removed as fast as they
are produces, and the surfaces of the polar caps consist mainly of
Fe$^{56}$.

In summary, the composition of the surface of the neutron star at its
polar caps is uncertain. Below we let the mass number, ${\rm A}$, of the
surface ions be a free parameter, and for numerical estimates we assume it
corresponds to Fe$^{56}$.

\subsection{2.3. The surface structure of neutron stars with
strong magnetic fields}

Whether an Arons-type model or a Ruderman-Sutherland-type model is
relevant for the polar-cap region of the pulsar magnetosphere depends
on whether or not charged particles can escape freely from the surface of
the neutron star due to thermionic emission. This in turn depends on
(a)~the binding energy, referred to here as the cohesive energy for ions
and as the work function for electrons, and (b)~the surface temperature.
The cohesive energy, $\Delta\varepsilon_c$, was overestimated in earlier
literature.

The structure of matter in the surface layers of neutron stars with
$B_\S\gg\alpha^2 B_{\rm cr}\simeq2.35\times10^5{\rm\,T}$,
where $\alpha=e^2/4\pi\varepsilon_0\hbar c=1/137$ is the fine structure
constant, is largely determined by the magnetic field  (Ruderman 1971;
cf.\ also Fushiki, Gudmundsson \and Pethick 1989 and
references therein). It has been suggested that the
surfaces of magnetic neutron stars with $B_\S\sim10^8\rm\,T$
consist of a phase of matter where atoms form chains aligned along the
field lines (Cheng, Ruderman \and Sutherland 1974; Flowers \et 1977).
The density of the magnetic metal phase is
$$\rho_\S\simeq
(4\times10^6{\rm\,kg\,m^{-3}})\left({B\over 10^8\,{\rm\,T}}
\right)^{6/5}\left({A\over56}\right)\left({Z\over26}\right)^{-3/5}\,.
\eqno(2.4)$$

\noindent The cohesive energy of the condensed Fe$^{56}$ matter was
estimated by Flowers \et (1977), using a variational method, to be
$$\Delta\varepsilon_c\simeq
(2.6{\rm\,keV})\left({B\over10^8{\rm\,T}}
\right)^{0.7}\,.
\eqno(2.5)$$

\noindent The properties of both bulk matter and isolated atoms in
strong magnetic fields have been reconsidered using various
different methods, including variational methods (M\"uller 1984;
Skjervold \and \"Ostgaard 1984b), the Thomas-Fermi and related methods
(Skjervold \and \"Ostgaard 1984a), density functional methods (Jones
1985, 1986), the Hartree-Fork method (Neuhauser, Langanke \and Koonin
1986; Neuhauser, Koonin \and Langanke 1987), and the Thomas-Fermi-Dirac
method with the Weizs\"acker gradient correction
(Abrahams \and Shapiro 1991). These investigations imply that
Flowers \et (1977) overestimated
the cohesive energy by a factor of at least a few. The most recent
Thomas-Fermi-Dirac calculations with the Weizs\"acker
coefficient $\lambda=1/9$ (the TFD-$1\over9$W method) gives
cohesive energy for Fe$^{56}$ to be
$\Delta\varepsilon_c\simeq0.91\rm\,keV$ for $B=10^8\rm\,T$,
$\Delta\varepsilon_c\simeq2.9\rm\,keV$ for $B=5\times10^8\rm\,T$,
and $\Delta\varepsilon_c\simeq4.9\rm\,keV$ for $B=10^9{\rm\,T}$
(Abrahams \and Shapiro 1991). The results of these calculations may be
approximated by
$$\Delta \varepsilon_c\simeq
(0.9{\rm\,keV})\left({B\over10^8{\rm\,T}}\right)^{0.73}\,.
\eqno(2.6)$$

\noindent The errors introduced by the TFD-${1\over9}$W method are
near one per cent of the binding energy of isolated atoms, which is
(Abrahams \and Shapiro 1991)
$$\varepsilon_b\simeq-
(55{\rm\,keV})\left({Z\over 26}\right)^{9/5}
\left({B\over10^8\rm\,T}\right)^{2/5}\,.
\eqno(2.7)$$

\noindent From (2.6) and (2.7), it follows that at $B\sim10^8\rm\,T$
the errors of the calculations, $\sim10^{-2}\varepsilon_b$, are of the
same order as the cohesive energy. Thus, the qualitative
result of Flowers \et (1977) that chains are energetically favored over
individual atoms is questionable. It is worth noting that the value of
$\lambda=1/9$ is the standard gradient expansion result for the
nonmagnetic case (Lieb 1981 and references therein). If  the value of
$\lambda$ for strong magnetic fields is smaller than $1/9$, the cohesive
energy is smaller than implied by (2.6). Moreover, if $\lambda$ is small
enough, free atoms are preferred over chains for $Z=26$ at $B\la$ a few
$\times10^8\rm\,T$ (Neuhauser \et 1986; Neuhauser \et 1987). Hence,
at $B_\S\la$ a few $\times10^8\rm\,T$ there are two possibilities:
either Fe$^{56}$ does not form a magnetic metal at the neutron star
surface, or the surface is solid with a small cohesive energy per Fe atom,
$\Delta\varepsilon_c<2$ -- $3\rm\,keV$ (Neuhauser \et 1986). The
existence of a magnetic metal is then unimportant for ejection of
particles from the surface, as discussed below.

\par For $B>(0.5$ -- $1)\times10^9{\rm\,T}$
the surface of a neutron star is probably
a magnetic metal, provided that the
surface temperature is smaller than the melting temperature,
$T_m$. An estimate of $T_m$ is (Slattery,
Doolen \and DeWitt 1980; Shapiro \and Teukolsky 1983; Ogata \and
Ichimaru 1990)
$$T_m\simeq
(5\times10^6{\rm\,K})\left({Z\over26}\right)^2
\left({A\over56} \right)^{-1/3}
\left({\rho \over10^7\,{\rm kg\, m}^{-3}}\right)^{1/3}\,,
\eqno(2.8)$$

\noindent where $\rho$, the density of the magnetic metal, may differ from
$\rho_\S$, as given by (2.4), by a factor two or so (Abrahams \and Shapiro
1991). The cohesive energy in such a strong magnetic field may be more
than $(3$ -- $5)\rm\,keV$.

\subsection{2.4.  Electron ejection {\rm(${\bf\Omega\cdot B}>0$)}}

The flow of electrons away from the surface of a pulsar with
${\bf\Omega\cdot B}>0$, assuming the neutron star surface to be a magnetic
metal, is determined either by thermionic emission or by field emission.

The important condition is that the current density, $J_{th}$, from
thermionic emission provide the Goldreich-Julian charge density. The
particles are quickly accelerated to relativistic energies above the
surface, and hence this condition becomes
$$J_{th} > n_\GJ ec\,,
\eqno(2.9)$$

\noindent Screening due to this charge density then implies
$E_\parallel=0$ on the stellar surface.

For electrons, the current density due to thermionic emission
is determined by the Richardson-Dushman equation (e.g., Gopal 1974):
$$J_{th} ={em\over2\pi^2\hbar^3}(kT)^2
\exp\left(-{w\over kT}\right)\,,
\eqno(2.10)$$

\noindent where $m$ is the electron mass, $k$ is the Boltzmann constant
and $w$ is the work function of electrons. Typically, the value of $w$ is
near the Fermi energy, which is given for a magnetic metal by (Ruderman
1971; Flowers \et 1977)
$$\varepsilon_\F={2\pi^4\hbar^4Z^2\rho^2\over
e^2m_p^2mA^2B^2}\,.
\eqno(2.11)$$

\noindent From (2.4) and (2.11), we have
$$\varepsilon_\F\simeq(0.8{\rm\,keV})
\left({Z\over 26}\right)^{4/5}
\left({B\over 10^8{\rm\,T}}\right)^{2/5}\,.
\eqno(2.12)$$

\noindent Taking $w\simeq\varepsilon_\F$ and using
(2.2), (2.10) and (2.12), we can rewrite the condition (2.9) in terms of
the characteristic temperature, $T_e$, identified as
$$T_e=0.04{w\over k}\simeq
(3.7\times10^5{\rm\,K})
\left({Z\over26}\right)^{4/5}
\left({B_\S\over10^8\rm\,T}\right)^{2/5}.
\eqno(2.13)$$

\noindent Then setting $T=T_\S$ as the surface temperature, (2.9) requires
$T_\S>T_e$ for thermionic emission of electrons to be adequate to
provide the Goldreich-Julian density.

Field emission is relevant only if thermionic emission is inadequate,
that is, for $T_\S<T_e$. Then $E_\parallel$ is nonzero on the surface
at the polar caps and the ejection of electrons results from quantum
mechanical tunneling through the barrier provided by the work
function and $E_\parallel$. The current density due to this
tunneling is (Beskin 1982)
$J=ME_\parallel\exp(-N/E_\parallel)$,
$M=3\times10^{16}(B/10^8\rm\,T)
(w/1{\rm\,keV})^{-1/2}\rm\,s^{-1}$,
$N=2\times10^{14}(w/1{\rm\,keV})^{3/2}\rm\,V\,m^{-1}$. Equating this
to the current implied by an electron density $n_\GJ$ outflowing at
the speed of light allows one to define a characteristic electric field:
$$E_e\simeq
(6\times10^{12}{\rm V\,m^{-1}})
\left({w\over 1\,{\rm keV}}\right)^{3/2}\,,
\eqno(2.14)$$

\noindent such that one requires $E_\parallel>E_e$ for
field emission to be effective.
For $E_\parallel\ll E_e$, the density of electrons ejected due to
field emission is negligible. For $E_\parallel\ga E_e$ the electrons
escape freely until their density of electrons builds up to $\sim
n_\GJ$, and they then screen the electric field, restricting it to
$E_\parallel\sim E_e$.

\subsection{2.5 Ion ejection {\rm(${\bf\Omega\cdot B}<0$)}}

The flow of ions away from the solid surface of a pulsar with
${\bf\Omega\cdot B}<0$ is limited by the rate of thermionic emission
of ions. The density of outflowing ions is (Cheng \and Ruderman 1980)
$$n_i\simeq n_\GJ\,{\rm min}(1,\zeta )\,,
\eqno(2.15)$$

\noindent with
$$\zeta=
\left({kT_\S\over 1\,{\rm keV}}\right)^{1/2}
\left({P\over 1\,{\rm s}}\right)
\left({B_\S\over 10^8{\rm\,T}}\right)^{-1}
\left({Z\over 26}\right)
\left({A\over 56}\right)^{-3/2}
\left({\rho_\S\over 10^7\,{\rm kg\,m}^{-3}}\right)
\exp
\left({-{30\Delta\varepsilon_c\over kT_\S}}
\right)\,.
\eqno(2.16)$$

\noindent The value of $n_i$ is very sensitive to the surface temperature
for $T_\S\sim T_i$, where $T_i$ is the characteristic temperature
$$T_i\simeq
{\Delta\varepsilon_c\over30{\rm\,k}}
\simeq(3.5\times10^5{\rm\,K})
\left({B_\S\over10^8{\rm\,T}}\right)^{0.73}\,,
\eqno(2.17)$$

\noindent where the cohesive energy of the condensed Fe$^{56}$
matter in a strong magnetic field is estimated from (2.6).
A small change in $T_\S$ around
$T_\S\sim T_i$ can have a large effect on $n_i$, with a change by a factor
of two causing $n_i/n_\GJ$ to vary from exponentially small to
approximately unity. For $T_\S>T_i$ the existence of a magnetic metal
at the neutron star surface does not affect the ejection of ions,
and one has $E_\parallel=0$ on the stellar surface.

\par Field emission of Fe$^{56}$ ions from a cold magnetic metal
at the surface of pulsars with $B_\S>10^8\rm\,T$
is unimportant for all known pulsars (Ginzburg \and Usov 1972).
Therefore, in the case $T_\S<T_i$ there is no ejection of
ions from the neutron star surface, and the electric field
component $E_\parallel$ near the polar caps
is nonzero and is determined by the polar-gap structure (e.g., Ruderman
\and Sutherland 1975 and below).

\section{3. PARTICLE ACCELERATION AND CURVATURE RADIATION}

The acceleration of primary particles is due to the parallel electric
field in the polar gaps. Curvature radiation is an important ingredient in
the generation of the secondary pair plasma, and it can also limit the
energy of the primary particles.

\subsection{3.1. Particle acceleration and curvature radiation}

If the energy losses of a particle are neglected, then a primary particle
ejected at the stellar surface reaches a maximum energy
$$\varepsilon=\Gamma mc^2
=e\Delta \varphi
\eqno(3.1)$$

\noindent at the top edge of the polar gap. Here $\Gamma$ is the Lorentz
factor and $\Delta\varphi$ is the potential across the polar gap.

In a strong magnetic field near the pulsar surface, electrons
lose the momentum component transverse to the magnetic field very
rapidly (timescale $\ll R/c$) and move away from the pulsar along
the field lines. For such electrons in the ground-state Landau level,
any energy loss is negligible up to the Lorentz factors of $\sim 10$.
For $10\la \Gamma\la 10^2$, the energy loss due to cyclotron
resonant scattering of thermal X-rays from the neutron star surface
increases sharply and may be very important for both motion of electrons
and their radiation (Dermer 1990 and references therein). At very
high energies, $\Gamma \gg 10^3-10^4$, the main energy loss for
ultrarelativistic electrons in the pulsar magnetospheres is due
to curvature radiation. In this case the rate of energy loss is
(e.g., Ochelkov \and Usov 1980)

$$|{\dot\varepsilon}|=
{e^2c\over6\pi\varepsilon_0 R_{\rm c}^2}\Gamma^4
\eqno(3.2)$$

\noindent where $R_{\rm c}$ is the radius of curvature of the
magnetic field lines. Equation (3.1) is valid only if the rate of
energy loss, $|{\dot\varepsilon}|$, is smaller than the rate of
energy gain, $eE_\parallel$, due to acceleration by the parallel
electric field.  The maximum energy of electrons cannot exceed the value
implied by balancing these energy gains and losses, that is, the maximum
is determined by $|{\dot\varepsilon}| \simeq eE_\parallel c$. The Lorentz
factor of particles when this quasi-stationary state is achieved is
$$\Gamma\simeq
\Gamma_{\rm st}=
\left(
{6\pi\varepsilon_0E_\parallel R_{\rm c}^2\over e}\right)^{1/4}\,.
\eqno(3.3)$$

\par Electrons in a strong magnetic field quickly radiate away their
perpendicular momentum and are then in their lowest Landau level. Such
electrons move along curved magnetic field lines and emit curvature
radiation. The mean energy, $\bar \varepsilon_\gamma$, of curvature
photons generated by ultrarelativistic electrons, $\Gamma \gg 1$, is
$$\bar\varepsilon_\gamma=
{3\over2}{\hbar c\over R_{\rm c}}\Gamma^3\,.
\eqno(3.4)$$

\noindent For $\Delta\varphi=\Delta\varphi_\RS$,
$\Gamma=e\Delta\varphi_\RS/mc^2\simeq 3\times10^6$ (Ruderman \and
Sutherland 1975) and $R_{\rm c}\simeq10R\simeq10^5\rm\,m$, from (3.4) we
have $\bar\varepsilon_\gamma\simeq10^2\rm\,MeV$. Thus the curvature
radiation of electrons accelerated in the polar gaps falls in
the $\gamma$-ray range. Note also that the mean energy satisfies
$\bar\varepsilon_\gamma\gg2mc^2\simeq1\rm\,MeV$.

\subsection{3.2. Curvature $\gamma$ rays and pair creation threshold}

A photon propagating in a strong magnetic field may decay into two
photons or into pairs (Toll 1952; Klepikov 1954; Erber 1966; Adler 1971
and section~4 below). Pair creation occurs only when the energy of the
photons exceeds the threshold $\sim2mc^2$ in the frame in which the
photon is propagating perpendicular to the field lines (section 4.2 below).

Although a particle in its lowest Landau level has no perpendicular motion
in a uniform magnetic field, it does have a small perpendicular motion if
the field lines are curved. The perpendicular motion is at the curvature
drift velocity, ${\bf v}_c$, which is such that the Lorentz force,
$\propto{\bf v}_c\times{\bf B}$, causes the particle to follow the curved
field line (e.g., Zheleznyakov \and Shaposhnikov 1979). The curvature drift
implies a perpendicular momentum
$$p_{\rm cd}=
{p^2_\parallel\over
m\omega_\B R_{\rm c}}\,,
\eqno(3.5)$$

\noindent where $\omega_\B=eB/mc$ is the gyration frequency.
The angle, $\psi$, between the momentum vector ${\bf p}$
and the tangent to the magnetic field line, defined by
$p_{\rm cd}=p\sin\psi$, $p_\parallel=p\cos\psi$,
$p=\vert{\bf p}\vert=\Gamma mv$, is given by
$$\psi\simeq
{c\Gamma\over\omega_\B R_{\rm c}}\simeq
1.7\times10^{-8}
\left({B\over 10^8\,{\rm T}}\right)^{-1}
\left({\Gamma \over 10^8}\right)
\left({R_{\rm c}\over 10^5\,{\rm m}}\right)^{-1}\,,
\eqno(3.6)$$

\noindent where (3.5) is used for $\psi\ll1$.

\par A highly relativistic particle emits nearly all its radiation
in a cone with half angle $\sim\Gamma^{-1}\ll1$ about the momentum vector
${\bf p}$. Hence, the curvature photons are emitted in a cone nearly
parallel to the field lines, but centered on an angle $\psi$ away from the
field lines. The question arises as to whether the condition for decay
into pairs can be satisfied at the point of emission. The curvature
photons have a perpendicular energy
$\la\bar\varepsilon_\gamma(\psi+\Gamma^{-1})$. Using (3.4) and (3.6), this
energy is below the threshold ($2mc^2$) for pair creation for
$${3\over2}{c\Gamma^2\over\omega_\B R_{\rm c}}
\left({{c\Gamma^2\over\omega_\B
R_{\rm c}}+1}\right)
<{2mc^2\over\hbar\omega_\B}\,.
\eqno(3.7)$$

\noindent This condition corresponds to the Lorentz factor of
ultrarelativistic particles being smaller than
$$\Gamma_*=
\left\{{R_{\rm c}\omega_\B\over c}
\left[\left({{1\over4}
+{4\over3}{mc^2\over\hbar\omega_\B}}\right)^{1/2}
-{1\over2}\right]
\right\}^{1/2}\,.
\eqno(3.8)$$

\noindent In the limits of a weak and strong magnetic field one gets
$$\Gamma_*\simeq
\cases{
\left(
{\displaystyle{4R_{\rm c}^2m\omega_\B\over3\hbar}}
\right)^{1/4}
&\quad for $B\ll B_{\rm cr}$,
\cr
\noalign{\vskip3pt}
\left(
{\displaystyle{4R_{\rm c}mc\over3\hbar}}\right)^{1/2}
&\quad for
$B\gg B_{\rm cr}$,
\cr}
\eqno(3.9)$$

\noindent which gives numerical values
$$\Gamma_*\simeq
\left({R_{\rm c}\over 10^5\,{\rm m}}\right)^{1/2}
\cases{2.8\times 10^8
\left(
{\displaystyle{B\over10^8\,{\rm T}}}
\right)^{1/4}
&\quad for $B\ll B_{\rm cr}$,
\cr
\noalign{\vskip3pt}
5.8\times10^8
&\quad for
$B\gg B_{\rm cr}$.
\cr}
\eqno(3.10)$$

\par Inside the polar gap, the value of $\Gamma$ is restricted by
$\Gamma\la\Gamma_{\rm st}$, where $\Gamma_{\rm st}$ depends on
$E_\parallel$, cf.\  (3.3).
The upper limit on $E_\parallel$ when charged particles flow freely from
the neutron star surface is (Arons 1981)
$$E^\A_\parallel\simeq
{1\over 8\sqrt{3}}
\left({\Omega R\over c}\right)^{5/2}cB_\S\,,
\eqno(3.11)$$

\noindent  and when there is no particle flow from the surface the upper
limit is (Ruderman \and Sutherland 1975)
$$E^\RS_\parallel\simeq \left({\Omega R\over c}
\right)^{3/2}cB_\S\,.
\eqno(3.12)$$

\noindent For all known pulsars one has $E^\RS_\parallel\sim(10^2$ --
$10^4)\,E^\A_\parallel$, implying $E_\parallel^\RS\gg E_\parallel^\A$.
\par Substituting $E^\RS_\parallel$ into
(3.3) and using (3.9) we have
$${\Gamma \over \Gamma_*}<
{\Gamma_{\rm st}\over\Gamma_*}<
0.2\left({P\over 0.1\,{\rm s}}\right)^{-3/8}
\cases{
1
&\quad for $B{_\S}\ll B_{\rm cr}$,
\cr
\noalign{\vskip3pt}
\left(
{\displaystyle{B_\S\over B_{\rm cr}}}\right)^{1/4}
&\quad for
$B{_\S}\gg B_{\rm cr}$,
\cr}
\eqno(3.13)$$

\noindent From (3.13) it follows that $\Gamma$ is smaller
than $\Gamma_*$ for all known pulsars except the millisecond
pulsars. However, the magnetic field at the surface of millisecond
pulsars, $B_\S\sim10^5\rm\,T$, is not strong enough to determine the
structure of matter, and the magnetic metal phase does not form in
the surface layers of the neutron stars (section~2).
Therefore, particles flow freely from the surface of such a
pulsar. In this case, $E^\A_\parallel$ is an
upper limit on $E_\parallel$ in the pulsar magnetospheres.
Substituting $E^\A_\parallel$ into (3.3) for $E_\parallel$ we get
$${\Gamma\over \Gamma_*}<
{\Gamma_{\rm st}\over\Gamma_*}<
2\times 10^{-2} \left({P\over 0.1\,{\rm s}}\right)^{-5/8}\,.
\eqno(3.14)$$

\noindent Then the condition $P>10^{-3}\rm\,s$, valid for all pulsars,
in (3.14) implies $\Gamma<0.3\Gamma_*$ for the millisecond pulsars.

It follows that for all known pulsars the curvature photons
generated near the neutron star surface are produced in a state
below the pair creation threshold. In order to reach the threshold
for decay into pairs (or the formation of positronium) the curvature
photons must travel some distance so that the angle between their
wave vector and the magnetic field has increased sufficiently.

\section{4. PROPAGATION OF $\gamma$ RAYS AND PAIR CREATION}

The propagation of $\gamma$ rays in a strong magnetic field
can lead both to splitting of one photon into two or more photons and to
decay of a photon into a free or bound electron-positron pair. These
processes can depend on which wave mode the photon is in.

\subsection{4.1. The modes of propagation for $\gamma$ rays}

\par The conventional expression for the refractive index of a plasma,
with the vacuum polarization by the magnetic field taken into account,
differs from unity by the order of
$\lbrack0.1\alpha(B/B_{\rm cr})^2
+(\omega_p/\omega)^2\rbrack
\sin^2\vartheta$, with
$\omega_p=(e^2n_p/\varepsilon_0m)^{1/2}$,
where $n_p$ is the plasma density, $\omega=\varepsilon_\gamma/\hbar$ is the
photon frequency and $\vartheta$ is the angle between the photon wave
vector ${\bf K}$ and the magnetic field ${\bf B}$ (Erber 1966; Adler 1971;
Shabad 1975). For $\omega\gg3\alpha^{-1/2}(B_{\rm cr}/B)\omega_p$, the
vacuum polarization gives the main contribution to the difference between
the refractive index and unity. Near the pulsar surface, where
$B\ga10^8\rm\,T$ and $n_p\la10^{26}\rm\,m^{-3}$ (Sturrock 1971;
Ruderman \and Sutherland 1975; Arons 1979, 1981, 1983), this condition
becomes $\omega\gg10^{18}\rm\,s^{-1}$, which is well satisfied for
$\gamma$-quanta. Hence, to understand the process of $\gamma$-quantum
propagation in the vicinity of a pulsar with $B_\S>10^8\rm\,T$, it
suffices to consider propagation in the vacuum polarized by a strong
magnetic field.

\par The principal modes of propagation for a photon in the magnetized
vacuum are linearly polarized with electric vectors either
perpendicular ($\perp$ mode) or parallel ($\parallel$ mode) to the
plane formed by the photon wave vector ${\bf K}$ and the vector
${\bf B}$. The labeling convention adopted here is standard,
although other labelings of the the modes are
used (e.g., Adler 1971; Usov \and Shabad 1983).
Both modes are generated in the process of the curvature radiation
(Jackson 1975).

Photons in either mode can decay into a free electron-positron pair
provided that a relevant threshold is exceeded (Toll 1952).
For the $\parallel$ mode both the electron and the positron can
be in their ground states, and the relevant threshold is
$$\varepsilon_\gamma\sin\vartheta=2mc^2\,.
\eqno(4.1)$$

\noindent For the $\perp$ mode the lowest allowed state requires that
either the electron or the positron being in the first excited state,
with the other in its ground state. The relevant threshold is then
$$\varepsilon_\gamma\sin\vartheta=
mc^2\lbrace1+\lbrack1+(2B/B_{\rm cr})\rbrack^{1/2}\rbrace\,.
\eqno(4.2)$$

\subsection{4.2. The curvature photon splitting in a strong
magnetic field}

While the photon is below the pair creation threshold,
its main (inelastic) interaction with the magnetic field is a
splitting into two photons $\gamma + B \rightarrow \gamma'+\gamma''+ B$
(e.g.,  Adler \et 1970; Bialynicka-Birula \and Bialynicki-Birula 1970;
Adler 1971; Stoneham 1979; Melrose 1983; Usov \and Shabad 1983; Baring 1991
and
references therein). For $B\la B_{\rm cr}$, when radiative corrections are
negligible, Bialynicka-Birula \and Bialynicki-Birula (1970)
showed that the splitting of a photon into more than two photons
is unimportant. Adler (1971) showed that
splitting involving an odd number of  $\parallel$-polarized photons
($\parallel\to \parallel+\parallel$, $\parallel\to\perp+\perp$,
$\perp\to\parallel+\perp$) is forbidden (by CP invariance), and that
below the pair creation threshold, of the remaining transitions the only
one that is kinematically allowed is  $\perp\to \parallel+\parallel$.
The probability for $\perp$-polarized photon splitting in the weak-field
limit, $B\la B_{\rm cr}$, is greatest when the energy of the initial
photons is divided equally between the two final photons.

\par The optical depth $\tau_d$ for the splitting of $\perp$-polarized
photons with $\varepsilon_\gamma\gg mc^2$ as they propagate from
$\vartheta \simeq 0$ up to the pair creation threshold (4.2) is (Usov \and
Shabad 1983)
$$\tau_d\simeq 1.7\times 10^5\left({B\over B_{\rm
cr}}\right)^6 \left[ 1+\left( 1+ {2B\over B_{\rm cr}}\right)^{1/2}
\right]^7
\left({R_{\rm c}\over 10^5\,{\rm m}}\right)
\left({mc^2\over \varepsilon_\gamma}
\right)^2\,.
\eqno(4.3)$$

\noindent It follows from (4.3) that if the strength of the magnetic
field at the pulsar poles is high enough, $B_\S\ga10^9\rm\,T$, most of
the $\perp$-polarized photons with $\varepsilon_\gamma\la10^2\rm\,MeV$
produced by curvature mechanism near the pulsar surface, are split
and transformed into the $\parallel$-polarized photons before
the pair creation threshold is reached. However, the $\parallel$-polarized
photons cannot be split below the pair creation threshold.

\subsection{4.3. Bound pair production in a strong magnetic field}

If the photon energy is above the pair creation threshold, the main process
by which a photon interacts with the magnetic field is single-photon
absorption, accompanied by pair creation:
$\gamma+B\rightarrow e^++e^-+B$ (Klepikov 1954; Erber 1966; Tsai \and Erber
1974; Melrose \and Parle 1983; Shabad \and Usov 1984). In the application
to
pulsars it is usually assumed that the curvature $\gamma$-quanta
produced below the pair creation threshold propagate as polarized photons
through the pulsar magnetosphere until they are absorbed by creating free
pairs. However, before a curvature photon reaches the threshold for free
pair creation it must cross the threshold for bound pair creation (Usov
\and Shabad 1985; Shabad \and Usov 1985, 1986, cf.\ also Pavlov \and
M\'esz\'aros 1993 and references therein). Hence, when considering the
creation of free pairs by curvature photons in the strong magnetic fields
of pulsars, the process of conversion into bound pairs, that is, into
positronium atoms, needs to be considered.

\subsection{4.3.1. Kinematic condition for positronium production}

To specify the positronium state created by a photon
with a given momentum $\hbar{\bf K}$, we choose a set of quantum
numbers suggested by the special choice of gauge with
scalar potential zero and vector potential with components
$A_x=-By$, $A_y=A_z=0$ for the constant and homogeneous magnetic
field $B=B_z$, $B_x=B_y=0$. Since ${\bf A}$ does not depend on $x$, $z$ or
$t$, the $x$- and $z$-components of momentum and the energy are constants
of the motion. Without loss of generality we choose the $x$-axis along
$\hbar{\bf K}_\perp$. Conservation of the $z$-component of momentum
implies that the parallel momentum of the center of mass of the
positronium is equal to the parallel component of the photon momentum:
$\hbar K_\parallel=P_z=p^+_z +p^-_z$, where $p_z^\pm$ are the parallel
momentum components of the electron and positron.

For present purposes it suffices to consider the case
$B\gg\alpha^2B_{\rm cr}\simeq2.35\times10^5\rm\,T$, in which
the orbital radius of the electron, $r_\B=(\hbar/eB)^{1/2}$, is
much less than the Bohr radius, $a=4\pi\varepsilon_0\hbar^2/me^2$. In this
case the dependence of the electron and positron wave functions on the
perpendicular motion is of the same form as when they do not form
positronium (Schiff \and Snyder 1939). The perpendicular motions of the
electron and positron are then each described by a continuous and a
discrete quantum number. The continuous quantum numbers, $p^\pm_x$, are
interpreted as the $y$-coordinates of the gyrocenters $y^\pm$ through
$$y^\pm=\mp{p^\pm\over eB}\,.
\eqno(4.4)$$

\noindent Conservation of the $x$-component of momentum then
implies $\hbar K_\perp=P_x=p^+_x+p^-_x$. Thus, a photon with
perpendicular momentum component $\hbar K_\perp$ produces an electron and
a positron whose gyrocenters are separated along the $y$-axis by
$\hbar K_\perp/eB$. The Coulomb interaction between the electron and
the positron affects only the wave functions of the relative motion
along the magnetic field, and reduces to the problem of the
one-dimensional hydrogen atom (Loudon 1959).

Thus, in a strong magnetic field the energy states of positronium in
its rest frame may be labeled by five quantum numbers: $n\ge0$ and
$n'\ge0$ for the Landau levels of the electron and positron,
$n_c\ge0$ for the principal quantum number of the one-dimensional
hydrogen-like atom, the parity of the hydrogen-like atom state, and
$P^2_x\ge0$ for the separation of gyrocenters of the electron and
positron. The energy of positronium does not depend on
$P_x^c=p^+_x-p^-_x$, reflecting the fact that the energy cannot depend on
the location of the center of mass of the positronium in a uniform
magnetic field.

The corrections to the dispersion relation for the two modes due to the
polarization of the vacuum are not important in the present context.
Hence we may approximate the dispersion relations, expressed as the
energy of a photon, by the vacuum relation
$$\varepsilon_\gamma=
\hbar c({K^2_\parallel+K^2_\perp})^{1/2}.
\eqno(4.5)$$

\noindent The energy of positronium may be written
$$\varepsilon_p=
\big[{P^2_zc^2+\varepsilon^2_{n n'}
(n_c,P_x^2\hbar^{-2})}\big]^{1/2},
\eqno(4.6)$$

\noindent where
$$\varepsilon_{n n'}(n_c,P^2_x\hbar^{-2})=mc^2
\left[
\left({1+{2nB\over B_{\rm cr}}}\right)^{1/2}
+\left({1+{2n'B\over B_{\rm cr}}}\right)^{1/2}
\right]
-\Delta\varepsilon_{n n'}(n_c,P^2_x\hbar^{-2})\,,
\eqno(4.7)$$

\noindent is the rest energy of the positronium, with
$\Delta\varepsilon_{n n'}(n_c,P^2_x\hbar^{-2})$ its binding
energy.

\par Using both momentum conservation,
$\hbar K_\perp=P_x$ and $\hbar K_\parallel=P_z$, and energy
conservation, $\varepsilon_\gamma=\varepsilon_p$, we obtain the
kinematic condition for positronium production by a photon in the form of
an equation for $K_\perp$:
$$K_\perp={1\over c\hbar}\varepsilon_{n n'}
(n_c,K^2_\perp)\,.
\eqno(4.8)$$

\noindent In order to establish whether this equation admits any
solutions, one should know the positronium dispersion law, that is, the
energy dependence on $P_x$. Unlike the case of free pair production by a
photon in a magnetic field, to find the kinematic condition for bound pair
production a dynamic problem has to be solved. This problem was considered
by Shabad \and Usov (1986) in detail, and the energy of a positronium atom
was calculated for a state when all quantum numbers, $n$, $n'$, $n_c$,
$P_z$, $P_x$ and $P_x^c=p_x^+-p_x^-$, are arbitrary. Here the quantum
number $P_x^c =-eB(y^++y^-)$ describes
the location of the center of mass of the positronium in the
$y$-direction, cf.\ (4.4).

\subsection{4.3.2. Energy of positronium and mixed
photon-positronium states}

When the two photon dispersion curves (identical curves for each mode)
and the energy relations for positronium for each of the discrete states
are plotted in momentum space, the intersection points between the curves
define parameters where the photon and positronium states interact. The
interaction may be described in terms of the dispersion curves for the
photons and the energy curves for the positronium states reconnecting to
form a set of mixed states. Each of these mixed states is photon-like in
one limit and positronium-like in another limit. A kinematic restriction
implies that photons in the two different modes can interact only with a
positronium state of the appropriate parity. Hence, there are actually two
distinct sets of mixed states, each applying only to one photon mode and
one parity state of the positronium. An adiabatic change, corresponding to
evolution along the dispersion curve, then allows a photon in one mode to
evolve into positronium in the appropriate parity state.

Consider the simplest example, which is the ground state of the
positronium,
whose parity is positive and which interacts only with the $\parallel$
mode.
For the ground state, $n=n'=n_c=0$, the binding
energy is (Shabad \and Usov 1986)
$$\Delta\varepsilon_{00}(0,\,P^2_x\hbar^{-2})=
\alpha^2mc^2
\left[ \ln {a\over
r_\B\left({1+r_\B^2P^2_x\hbar^{-2}}\right)^{1/2} }\right]^2\,,
\eqno(4.9)$$

\noindent From this equation, we can see that the
binding energy is nonzero for any value of $P_x$ and
tends asymptotically to zero as $P_x^2$ tends to infinity.
Hence, if the magnetic field is strong enough, $B\gg \alpha^2B_{\rm cr}$,
the positronium energy $\varepsilon_{00}$ varies in a
narrow range from
$$\varepsilon_{00}(0,\,0)=mc^2
\lbrace2-\alpha^2\lbrack
\ln(a/r_\B)
\rbrack^2\rbrace\,,
\eqno(4.10)$$

\noindent to $2mc^2$ when $P_x^2$ varies from zero to infinity.
Therefore, (4.8) must have a solution. It follows that positronium
production by a photon in the $\parallel$ mode becomes kinematically
allowed before the threshold of free pair production (4.1) is
reached. A similar situation applies to a photon in
the $\perp$ mode near the second threshold of pair creation (Shabad
\and Usov 1986). Figure 1 solves (4.8) graphically for both
$\parallel$-polarized and $\perp$-polarized photons.

\par When the photon and positronium dispersion curves are
treated as independent, they are given by the following
equations (Figure 1):
$$\left({\varepsilon_\gamma\over\hbar c}\right)^2
-K^2_\parallel=
K^2_\perp\,,
\eqno(4.11)$$

\noindent for the photon, and
$$\left({\varepsilon_p\over\hbar c}\right)^2-K^2_\parallel=
\left({mc\over\hbar}\right)^2
\left\{2-\alpha^2
\left[
\ln{a\over r_\B\left({1+r_\B^2K^2_\perp}\right)^{1/2}}
\right]^2
\right\}^2
\eqno(4.12)$$

\noindent for the positronium with $n=n'=n_c=0$. In (4.12)
the parallel and perpendicular wave vectors,
$K_\parallel=P_z\hbar^{-1}$ and
$K_\perp=P_x\hbar^{-1}$, are used instead of $P_z$ and
$P_x$ to describe the positronium state. Mutual
transformations of the photon and positronium are kinematically
allowed at the point satisfying $K_\perp=(K_\perp)_0$,
$$(K_\perp)_0\simeq
{mc\over\hbar}
\left\{
{4-\alpha^2
\left(
\ln{a\over r_\B[{1+(4B_{\rm cr}/B)}]^{1/2}}\right)^2}
\right\}^{1/2},
\eqno(4.13)$$

\noindent where, in the absence of any interference, the photon and
positronium dispersion curves would intersect.

\par The dispersion curves of the photon and the positronium
interfere strongly near $K^2_\perp=(K_\perp)_0^2$. The dispersion
curves repulse each other and reconnect to form mixed states (Figure
2). This effect is quite general (Von Neumann \and Wigner 1964) and is
sometimes called the theorem on nonintersection of spectral terms.

\par For $n=n'=n_c=0$, the split dispersion curves of
the mixed states are (Shabad \and Usov 1985, 1986)
$$\left({\varepsilon\over\hbar c}\right)^2-K^2_\parallel=
{1\over2}
\left(
\left[
{\varepsilon_{00}(0,K^2_\perp)
\over\hbar c}
\right]^2
+K^2_\perp\pm
\left\{
{\left[\left(
{\varepsilon_{00}(0,K^2_\perp)
\over\hbar c}\right)^2
-K^2_\perp\right]^2
+4A(K^2_\perp)}
\right\}^{1/2}
\right),
\eqno(4.14)$$

\noindent with
$$A(K^2_\perp)=
{4\alpha eB\varepsilon_{00}(0,K^2_\perp)\over\hbar^2c^2a}\,
\ln\left[
{a\over r_\B
({1+r_\B^2K^2_\perp})^{1/2}}
\right]\,
\exp\left(-{\hbar K^2_\perp\over2eB}\right)\,.
\eqno(4.15)$$

\noindent The lower of the two dispersion curves (the minus sign in
(4.14)) is photon-like for $K^2_\perp\ll(K_\perp)_0^2$, and
positronium-like for $K^2_\perp\gg(K_\perp)_0^2$ (Figure 2). The upper
branch in Figure 2 is positronium-like
for $K^2_\perp\ll(K_\perp)_0^2$. This upper branch crosses the
dispersion curve of the odd state of the positronium with $n_c=1$
without interfering with it (Shabad \and Usov 1986). As
$K^2_\perp$ grows further, the upper branch approaches the even
positronium state with $n_c=1$. There is an infinite set of bound
positronium states at higher quantum numbers, and hence an infinite number
of branches of the mixed-state dispersion curves below the limit,
$(\varepsilon/\hbar c)^2-K^2_\parallel=(2mc/\hbar)^2$, that separates
the bound states from the continuum of unbound electron-positron states.

\subsection{4.3.3. Channelling along the magnetic field lines}

A curvature photon is emitted almost parallel to the magnetic
field below the pair creation threshold, $K^2_\perp\ll(2mc/\hbar)^2$
(section~3.3). Provided that the geometrical optics or adiabatic
approximation applies, the curvature of the magnetic field lines cause the
photon to shift along the lower branch of the dispersion curves in Figure
2. As a result it passes through the mixed photon-positronium state and
gradually turns into positronium when the condition
$$K^2_\perp>
(2mc^2/\hbar)^2+\Delta K^2_\perp
\eqno(4.16)$$

\noindent is satisfied, where
$$\Delta K^2_\perp=
2\big[{A((2mc/\hbar)^2)}\big]^{1/2}=
4\alpha\left({mc\over\hbar}\right)^2
\left\{{2B\over B_{\rm cr}}
\ln{a\over r_\B[{1+(4B_{\rm cr}/B)}]^{1/2}}
\right\}^{1/2}
\exp\left(-{B_{\rm cr}\over B}\right)
\eqno(4.17)$$

\noindent is a measure of the gap between the lower and upper branches.

In order for $\gamma$-quantum absorption to lead to the formation of
bound pairs rather than free pairs, the adiabatic approximation must
apply. This requires that the time (Shabad \and Usov 1986)
$$\tau_c\simeq
{\varepsilon_\gamma\over4mc^2}
{\Delta K^2_\perp\over K^2_\parallel}{R_{\rm c}\over c}
\eqno(4.18)$$

\noindent needed to convert from the photon-dominated into
positronium-dominated state should be much longer than the
inverse frequency of the photon, $\omega^{-1}=\hbar/\varepsilon_\gamma$.
For photon energies $\varepsilon_\gamma\gg2mc^2$ which are of interest
here, we have $K^2_\parallel\simeq(\varepsilon_\gamma/\hbar c)^2$.
Taking this into account and using (4.15) -- (4.17), the condition for the
adiabatic approximation to apply, $\tau_c\gg\omega^{-1}$, can be written
as a restriction on the magnetic field:
$$\left[{2B\over B_{\rm cr}}
\ln{a\over r_\B[{1+(4B_{\rm cr}/B)}]^{1/2}}
\right]^{1/2}
\exp\left(-{B_{\rm cr}\over B}\right)\gg
{a\over R_{\rm c}}\,.
\eqno(4.19)$$

\noindent For $R_{\rm c}\sim10R\simeq10^5\rm\,m$, (4.19) yields
$B>0.03\,B_{\rm cr}$.

\par Another restriction is imposed by a widening of the dispersion
curves that occurs because of the finite lifetimes of all excited
states. In the case of the $\parallel$-polarized photons, for kinematic
reasons the state described by the lower dispersion curve of equation
(4.14) cannot decay into two photons (cf.\ section~4.2 and references
therein). However, the positronium-dominated state which is described by
the left part of the upper dispersion curve of equation (4.14), may decay
into two photons. The probability of a two-photon positronium
annihilation in a strong magnetic field $(B\sim 10^{8}$ --
$10^9\rm\,T$) is (Wunner \and Herold 1979)
$$W_{2\gamma}\simeq
(8\times10^{12}{\rm\,s^{-1}})
\left({B\over10^8\rm\,T}
\right)\,,
\eqno(4.20)$$

\noindent for the case $K_\perp=0$ in the positronium rest frame. The
value of $W_{2\gamma}$ decreases as $K_\perp$ increases, so that
(4.20) gives the maximum value of $W_{2\gamma}$.

\par The widening of the dispersion curves can be neglected if the
energy gap between the $\pm$ branches is much larger than this widening.
{}From (4.14), in the rest frame of the photon-positronium state,
$K_\parallel=0$, the energy gap is (Shabad \and Usov 1986)
$$\Delta\varepsilon_g\simeq
\hbar^2\Delta K^2_\perp /4m\,.
\eqno(4.21)$$

\noindent The widening may be neglected in evaluating $\hbar W_{2\gamma}$.
The field which gives $\Delta\varepsilon_g /\hbar W_{2\gamma}=1$ may be
found from (4.17), (4.20) and (4.21) to be about $0.08\,B_{\rm cr}$. The
ratio $\Delta\varepsilon_g/\hbar W_{2\gamma}$ is sensitive to the
strength of the magnetic field: it is $\sim10^{-4}$ for
$B=0.05\,B_{\rm cr}$ and $\sim10$ for $B=0.1\,B_{\rm cr}$. Hence, for
$B\ga0.1\,B_{\rm cr}$ the conditions for applicability of the dispersion
curve (4.14) are fulfilled, and effectively all curvature
$\gamma$-quanta in the $\parallel$ mode are captured near the first
threshold of bound pair creation by gradually converting into
positronium. Positronium atoms with $n=n'=n_c=0$ and
$K^2_\perp>(K_\perp)_0^2$, are stable against spontaneous decay into
photons, which is forbidden by momentum and energy conservation (Herold
\et 1985; cf.\ also Adler 1971; Usov \and Shabad 1983). Thus the
positronium atoms formed by capture of the $\parallel$-polarized photons in
a strong magnetic field are stable, in the absence of such external
factors as electric fields and ionizing radiation.

\par The $\perp$-polarized photons cross the first threshold (4.1) without
any pair creation, and pair creation is allowed only when the threshold
(4.2) is reached. Positronium created by $\perp$-polarized
photons at this threshold have $n+n'\geq 1$, so that either the electron or
the positron is in an excited state with the spin quantum number,
$s=1$, opposite to that in the ground state. In this case the main reason
for the widening of the dispersion curve is a spin-flip transition,
$s=1\to s=-1$. The probability of such a transition, $W_s$, has been
calculated for both the non-relativistic (Daugherty \and Ventura 1978;
Melrose \and Zheleznyakov 1981) and the relativistic (Herold, Ruder \and
Wunner 1982) cases. The energy gap between the dispersion curves for a
$\perp$-polarized photon mixed with the positronium state of the
second series, $n=n_c=0$, $n'=1$ or $n'=n_c=0$, $n=1$, differs little from
(4.21). The condition for bound pairs to dominate the absorption of
$\perp$-polarized photons in a strong magnetic field is
$\Delta\varepsilon_g>\hbar W_s$. Using the relativistic calculation of
$W_s$ (cf.\ Figure 2 in Herold \et 1982) and (4.21), we can write this
condition as a restriction on the strength of the magnetic field:
$B>0.15\,B_{\rm cr}$.
\par Summarizing, the main results of this section are as follows.
\smallskip
\item{ (1)} For $B<0.04\,B_{\rm cr}\simeq2\times10^8\rm\,T$, the
creation of free electron-positron pairs dominates.

\item{(2)} For $0.04\,B_{\rm cr}\la B<0.1\,B_{\rm cr}$,
the probabilities of creation of both positronium atoms and free
pairs are more or less comparable and a numerical treatment is
needed to determine the details.

\item{(3)} For $B\ga0.1$ -- $0.15\,B_{\rm cr}$ the curvature photons of
both modes are captured near the pair creation thresholds.
\smallskip\noindent
Note that when the condition $B>0.15\,B_{\rm cr}$ is satisfied
bound pair creation necessarily occurs. Bound pair creation occurs
for lower values of $B$, and dominates over free pair creation for
$B\ga0.1\,B_{\rm cr}$.

\subsection{4.4. The role of strong electric fields}

So far in our discussion of bound pair creation we
neglect the electric field in the pulsar
magnetosphere. In general, both parallel,
${\bf E}_\parallel$, and perpendicular, ${\bf E}_\perp$, components are
nonzero near the pulsar surface (Ruderman \and Sutherland 1975;
Arons 1979, 1981, 1983; Michel 1991).
The field ${\bf E}_\perp$ causes a drift of particles
across the magnetic field with velocity
$${\bf v}_\D={{\bf E}_\perp{\bf\times B}\over B^2}\,,
\eqno(4.22)$$

\noindent provided one has $E_\perp<B$, which condition is satisfied in the
cases of interest here. In the pulsar magnetosphere at a point with
radius vector ${\bf r}$, we may eliminate ${\bf E}_\perp$ by making a
Lorentz transformation, with ${\bf v}={\bf v}_\D({\bf r})$, to the
{\it drift frame}. Quantities in the drift frame are denoted by a tilde,
so that we have $\tilde {\bf E}_\perp=0$.

\par The characteristic time, $\tau_\D$, for a test particle
to be dragged into motion
across ${\bf B}$ with the mean velocity ${\bf v}_\D$ is of
the order of the time of one gyration:
$$\tau_\D\simeq\omega^{-1}_\B\Gamma
\simeq(10^{-12}{\rm\,s})
\left({B\over0.1\,B_{\rm cr}}\right)^{-1}
\left({\Gamma\over10^8}\right)\,.
\eqno(4.23)$$

\noindent For particles which are either ejected from the pulsar surface
or created in the pulsar magnetosphere, the initial value of the Lorentz
factor satisfies $\Gamma<\Gamma_{\rm st}\la$ a few $\times10^8$, cf.\
(3.3). As a consequence, the distance relativistic particles travel during
the time $\tau_\D$ is much smaller than the characteristic scale of the
electric field variation, of order the neutron star radius. Hence,
particles acquire the drift motion very rapidly.

\par If the drift velocity ${\bf v}_\D$ is nonuniform its effect cannot be
removed entirely by a Lorentz transformation. Suppose the drift velocity
varies along the field lines, so that the electric field $\tilde E_\perp$
changes with distance, $z$. Then, in addition to the curvature drift (3.6),
the polarization drift contributes to the angle $\tilde\psi$ between the
particle velocity $\tilde{\bf v}$ and the magnetic field $\tilde{\bf B}$
(Appendix A):
$$\tilde\psi\le
{m\Gamma\over eB^2}
{E_\perp\over\Delta l}
\simeq3.8\times10^{-8}
{E_\perp\over cB}
\left({\Delta l\over 10^4\,{\rm m}}\right)^{-1}
\left({B\over0.1\,B_{\rm cr}}\right)^{-1}
\left({\Gamma\over10^8}\right)\,,
\eqno(4.24)$$

\noindent which applies to a particle in its lowest Landau level, and
where $\Delta l$ is a characteristic scale length of the electric field
variation. Emission of curvature photons is centered on the direction
of the velocity of the particle, which is at angle $\tilde\psi$. For
these curvature photons to be produced below the pair creation threshold
requires
$$\tilde\varepsilon_\gamma\sin\tilde\psi<2mc^2\,.
\eqno(4.25)$$

\noindent Assuming $\tilde\Gamma\simeq\Gamma$,
$\tilde\varepsilon_\gamma\simeq\varepsilon_\gamma$,
$\tilde B\simeq B$, $\Gamma<\Gamma_{\rm st}$,
$E_\parallel\la(E^\RS_\parallel)_{\rm max}$
and $E_\perp/B\la\Omega R$, and using
(3.3), (3.4), (3.12) and (4.24), condition (4.25) gives
$${9\pi\varepsilon_0c\hbar\over2e^2}
\left({\Omega R\over c}\right)^{5/2}
{R_{\rm c}\over\Delta l}<1
\eqno(4.26)$$

\noindent or
$$P>(1.6\times10^{-3}{\rm\,s})
\left({R_{\rm c}\over \Delta l}\right)^{2/5}\,.
\eqno(4.27)$$

\noindent For $\Delta l\sim R_{\rm c}$ near the neutron star surface,
(4.27) is fulfilled for most pulsars with the only exceptions being a few
millisecond pulsars. However, as mentioned above,
the strength of the magnetic field at the surface of known
millisecond pulsars is $\sim10^5\rm\,T$, that is, much smaller than
$0.1\,B_{\rm cr}$. For all known pulsars with strong magnetic
fields, $B_\S\ga0.1\,B_{\rm cr}$, we have $P\gg10^{-3}\rm\,s$,
and hence the curvature photons generated near their surface are
produced in a state below the pair creation threshold
irrespective of existence of the field $E_\perp$ in
the pulsar magnetosphere.
Provided that the energy of the curvature photon is high enough,
$\varepsilon_\gamma\gg2mc^2$,
as the photon propagates it approaches
the threshold for pair creation.
With the characteristic scale of variation of the
field $E_\perp$ is many orders of magnitude larger than
the size of positronium atoms, at any instant
we can introduce the frame with ${\bf E}_\perp=0$.
In this frame our consideration of bound pair creation is
applicable, implying that bound pairs must be created by the
curvature photons for $B>0.1\,B_{\rm cr}$.

\par The parallel electric field can lead to {\it field ionization\/} of
the positronium. The Coulomb field and $E_\parallel$ combine to form a
potential barrier, and quantum mechanical tunneling through this barrier
leads to a nonzero probability for the decay of the bound state into a
free electron and a free positron (Herold \et 1985; Shabad \and Usov 1985;
Bhatia \et 1988). The probability of the field ionization in the rest
frame of the positronium is (Shabad \and Usov 1985)
$$W_\E=
{eE_\parallel\over2\sqrt{m}\,
(\Delta\varepsilon_{00})^{1/2}}\,\,
{\rm exp}\left[-4{\sqrt{m}\,
(\Delta\varepsilon_{00}})^{3/2}
\over3e\hbar E_\parallel\right]\,.
\eqno(4.28)$$

\noindent The value of $W_\E$ is very sensitive to the
field $E_\parallel$. Provided that the strength of the
field $E_\parallel$ near the pulsar is less than
$$E_\parallel^{\rm ion}
\simeq{2\over 3}{\sqrt{m}\,
(\Delta\varepsilon_{00})^{3/2}\over e\hbar}
\left(\ln{R\,\Delta\varepsilon_{00}\over c\hbar
\Gamma}\right)^{-1}\,,
\eqno(4.29)$$

\noindent the ionization probability of the positronium before it
escapes from the pulsar environment is negligible. For
$E_\parallel\ga2E_\parallel^{{\rm ion}}$,
the mean free path of positronium atoms is much smaller
than the neutron star radius, and positronium atoms are ionized
almost immediately after their formation. Moreover,
if the field $E_\parallel$ is so strong that the
characteristic time of the positronium ionization, $(W_\E)^{-1}$, is of
the order of or smaller than the characteristic time of the photon
conversion, $\tau_c$, into a bound pair, cf.\ (4.18) and (4.28), then
free electron-positron pairs rather than bound pairs are created
by the curvature photons.

\par Numerically, $E_\parallel^{\rm ion}$ varies slowly
with the magnetic field and with the Lorentz factor of the positronium.
For $B_\S\simeq(0.1$ -- $1)B_{\rm cr}$ and $\Gamma\sim10$ -- $10^6$,
(4.29) gives $E_\parallel^{\rm ion}\simeq10^{12}\,{\rm V\,m^{-1}}$ to
within a factor of 2 or so.

\subsection{4.5. The photoionization of positronium atoms}

Positronium atoms can be destroyed through ionization by photons. The main
source of such photons is thermal radiation from the neutron star
surface. The surface temperature of neutron stars, with ages $\sim10^3$ --
$10^6\rm\,yr$, is usually estimated to be $\sim10^5$ -- $10^6\rm\,K$
(e.g., Van Riper 1991; Page \and Applegate 1992 and references therein)
These estimates are more or less in agreement with recent observations of
X-rays from a few pulsars.
The mean free path for relativistic positronium, $\Gamma\gg1$,
against photoionization is
$$l_{\rm ph}\simeq
c\Gamma(W_{\rm ph})^{-1}\,,
\eqno(4.30)$$

\noindent where $W_{\rm ph}$ is the probability of the positronium
photoionization in the rest frame.

\par The value of $W_{\rm ph}$ was calculated numerically
by Bhatia \et (1992) for $B=4.7\times10^8\rm\,T$ and
$K_\perp=(K_\perp)_0\simeq2mc/\hbar$. The results of their calculations
may be approximated by the analytic expression
$$W_{\rm ph}\simeq
(6\times10^7{\rm\,s^{-1}})
\left({\Gamma\over10^2}\right)^{-2}
\left({T_\S\over10^6\,\,{\rm K}}\right)^2\,,
\eqno(4.31)$$

\noindent which is valid for $10^2\le\Gamma\le10^4$ and $10^5\le
T_\S\leq10^6\rm\,K$ with the accuracy $\sim 20$ per cent. Equations (4.30)
and (4.31) yield
$$l_{\rm ph}\simeq
(0.5\times10^3{\rm\,m})
\left({\Gamma\over10^2}\right)^3
\left({T_\S\over10^6\,{\rm K}}\right)^{-2}\,.
\eqno(4.32)$$

\noindent Note that only photons with energy $\varepsilon_\gamma\simeq(1$
-- $3)\Delta\varepsilon_{00}$ in the rest frame contribute to the decay of
the positronium (Bhatia \et 1992). Hence, the dependence of $W_{\rm ph}$
on $B$ and $K_\perp$ is through the $B$ and $K_\perp$ dependence of the
binding energy, which may be written in the form, cf.\ (4.9),
$$\Delta\varepsilon_{00}(0,K^2_\perp,B)\simeq
{\alpha^2\over4}mc^2
\left[\ln{B\over\alpha^2B_{\rm cr}}
\left(1+{4B_{\rm cr}\over B}
{K^2_\perp\over(2mc/\hbar)^2}\right)^{-1}
\right] ^2\,.
\eqno(4.33)$$

\noindent From (4.33) we infer that the $B$ and $K_\perp$ dependence of
$W_{\rm ph}$ is only logarithmic, and hence (4.32) may be used to estimate
the order of $l_{\rm ph}$ for $0.1\,B_{\rm cr}\la B\la B_{\rm cr}$ and
$K_\perp\ga2mc/\hbar$.

\section{5. POLAR GAPS AND NONTHERMAL LUMINOSITIES OF PULSARS}

As discussed in section~4, adiabatic conversion of the curvature
$\gamma$-quanta into mutually bound pairs in a strong magnetic field may
result in these pairs not screening the electric field near the pulsar.
The potential $\Delta\varphi$ across the polar gaps may then exceed
$\Delta\varphi_\RS$, leading to an increase in the theoretical
estimate of the total energy flux carried by relativistic particles
from the polar gaps into the pulsar magnetosphere. In this section
we explore how polar-gap models need to be modified to apply to
pulsars with strong magnetic fields to take account of the role
played by bound pairs. Assuming free ejections from the stellar surface,
we find that the modification gives at most a modest enhancement in the
energy flux in primary particles, and this is inadequate to account for
the observed nonthermal emission from some pulsars. Alternatively,
assuming limited ejection from the stellar surface, we argue that there is
a feedback mechanism that favors a self-consistent model. The energy flux
in primary particles in this model appears capable of accounting for the
observed nonthermal emission from most of the observed high-frequency
pulsars.

\subsection{5.1. Free ejection of electrons from the polar
cap}

First consider the case where electrons flow freely from the surface of
the neutron star, which occurs when the temperature of the polar caps
satisfies $T_\S>T_e$, cf.\ (2.13). The model of Arons is then assumed to
apply, with  $E_\parallel=0$ at the stellar surface.
In the outflowing plasma at distance $r$ from the center of the
neutron star, $E_\parallel$ may be as high as (Arons \and Scharlemann 1979)
$$E_{\parallel}\simeq E^\A_\parallel
\sin i \times\cases{
s/\Delta R_p & at $\quad 0 \leq s<\Delta R_p$,
\cr
(R/r)^{1/2}& at $\quad s>\Delta R_p$,
\cr}
\eqno(5.1)$$

\noindent where $i$ is the angle between the stellar magnetic moment
and the spin axis, $s=r-R$ is the distance from the surface of the
star, $\Delta R_p \simeq(\Omega R/c)^{1/2}R$ is the radius of the polar
cap, $\Delta R_p\ll R$, and $E_\parallel^\A$ is given by (3.11).
Here and below we consider a pulsar with a dipole magnetic field.
The potential which corresponds to the electric field
$E_\parallel$ is
$$\varphi(r)\simeq
{1\over8\sqrt3}
\left({\Omega R\over c}\right)^{5/2}
cB_\S\sin i \times
\cases{
{\displaystyle{s^2\over2\Delta R_p}}
&\quad at $s\leq\Delta R_p$,
\cr
\noalign{\vskip3pt}
2R{\displaystyle{
\left[\left({r\over R}\right)^{1/2}-1\right] }}
-{\displaystyle{\Delta R_p\over2}}
&\quad at $s>\Delta R_p$.
\cr}
\eqno(5.2)$$

\par The total power carried away by both relativistic particles and
radiation from the polar gap into the pulsar magnetosphere is
$$L_p\simeq
\dot N_{\rm prim}e\Delta\varphi \,,
\eqno(5.3)$$

\noindent where $\dot N_{\rm prim}$ is the flux of primary electrons from
the polar cap, $\Delta\varphi=\varphi(R+H)-\varphi(R)$ is the potential
across the polar gap and $H$ is the thickness of the polar gap.
Equation (5.3) is valid irrespective of whether the pairs created near the
pulsar are free or bound. The version of pair creation determines only the
thickness of the polar gap, $H$, and thereby affects
the value of $L_p$.

Assuming that the pairs created by the curvature photons are free,
the thickness of the polar gap, $H=H_f$, is (Arons 1981, 1983)
$$H_f\simeq\cases{(7\times10^2{\rm\,m})
\left(
{\displaystyle{B_\S\over10^8{\rm\,T}}}
\right)^{-2/5}
\left(
{\displaystyle{P\over 33\,\,{\rm ms}}}
\right)^{11/20}
(\sin i)^{-3/10}
&\quad at $P<P_d$,
\cr
\noalign{\vskip3pt}
(1\times10^4{\rm\,m})
\left(
{\displaystyle{B_\S\over10^8{\rm\,T}}}
\right)^{-1}
\left(
{\displaystyle{P\over0.1\,{\rm s}}}
\right)^{17/8}(\sin i)^{-3/4}
&\quad at $P>P_d$,
\cr}
\eqno(5.4)$$

\noindent where
$$P_d\simeq(0.03{\rm\,s})
\left({B_\S\over10^8{\rm\,T}}
\right)^{8/21}
(\sin i)^{2/7}
\eqno(5.5)$$

\noindent is the value of the pulsar period at which $H$ is equal to
$\Delta R_p$.

The formation of bound rather than free pairs is important for
$B_\S>0.1\,B_{\rm cr}$. Near the polar cap of the pulsar the condition
$B>0.1\,B_{\rm cr}$ is satisfied up to a height
$$H_b=R\left[
\left({B_\S\over0.1\,B_{\rm cr}}\right)^{1/3}
-1
\right]
\eqno(5.6)$$

\noindent above the pulsar surface. Thus, for $H_b>H_f$, bound pairs
rather than free pairs are created in the polar gaps. In this case,
provided that the probability of ionization of the bound pairs near the
polar caps is sufficiently small, the thickness of the polar gaps could
satisfy $H\sim H_b$. It follows that for $H_b>H_f$ one needs to consider
the ionization of bound pairs near the polar caps inside
the layer $0\leq s\leq H_b$ when estimating the values of both $H$
and $L^{\rm max}_{\rm nth}$.

\par For positronium atoms the probability of field ionization
is negligible for $E^\A_\parallel<E^{\rm ion}_\parallel
\simeq 10^{12}\,{\rm\,V\,m^{-1}}$ (section 4.4). Using (3.11)
this condition may be written in the form
$$P>(0.8\times10^{-2}{\rm\,s})
\left({B\over0.1\,B_{\rm cr}}\right)^{2/5}\,,
\eqno(5.7)$$

\noindent which is satisfied for all known pulsars.

\par The mean free path for photoionization increases rapidly with
increasing Lorentz factor of the bound pair, $l_{\rm ph}\propto\Gamma^3$,
cf.\ (4.32). It follows that the photoionization of bound pairs with
energy near the low-energy edge of their spectrum makes the main
contribution to the density of free particles inside the polar gaps. This
differs qualitatively from the case of $B<0.1\,B_{\rm cr}$ for which the
high-energy tail of curvature photons is responsible for both the creation
of free pairs in the polar gaps and the screening of the electric field
$E{_\parallel}$ (Ruderman \and Sutherland 1975; Arons 1981).

\par Let us estimate the height, $H_i$, above the stellar surface
at which photoionization of bound pairs is efficient enough to
determine the structure of the polar gaps. The lower limit on the Lorentz
factor of pairs created at the distance $s$ from the pulsar surface is
(Appendix B)
$$\Gamma_{\rm min}={4\over3}
\left({c\over \Omega R}\right)^{1/2}
{R\over s}\,.
\eqno(5.8)$$

\noindent The energy of the curvature photons responsible for the
creation of pairs with the Lorentz factor $\Gamma_{\rm min}$,
$\hbar\omega=2\Gamma_{\rm min}mc^2$, is less than the mean energy of the
curvature photons radiated by the primary particles in the polar gaps,
$\hbar\omega\ll(3\hbar c/2R_c)\Gamma^3$, where $\Gamma$ is the
Lorentz factor of primary particles accelerated in the polar gaps (section
3.1). In this case the spectral power of curvature radiation does not
depend on $\Gamma$ and is given by (e.g., Ochelkov \and Usov 1980)
$$p(\omega)\simeq
{e^2\over4\pi\varepsilon_0 c}
\left({c\over R_c}\right)^{2/3}
\omega^{1/3}\,.
\eqno(5.9)$$

\noindent At distance $s$ from the pulsar surface
the density of the curvature photons with energy
$\hbar\omega=2\Gamma_{\rm min}mc^2$ is
$$n_\gamma\simeq
n_\GJ{p(\omega)\over\hbar}
{s\over c}\,.
\eqno(5.10)$$

Taking into account the motion of the pair, the probability of
the photoionization of bound pairs in the frame of the pulsar is
$W_{\rm ph}/\Gamma$. From (4.31) and (5.8) -- (5.10) we have the
following estimate of the density of free pairs, $n_\pm$, created
near the polar caps:
$$n_\pm\simeq n_\gamma
{W_{\rm ph}\over\Gamma_{\rm min}}
{s\over c}\,,
\eqno(5.11)$$

\noindent or
$${n_\pm\over n_\GJ}\simeq
1.6\times10^6
\left({P\over 0.1\,{\rm s}}\right)^{-4/3}
\left({s\over R}\right)^{14/3}
\left({T_\S\over10^6\,{\rm K}}\right)^2\,,
\eqno(5.12)$$

\noindent where we assume $R_{\rm c}=10^5\rm\,m$ in (5.9).
If at distance $s$ the density of free pairs, $n_\pm$,
exceeds about $n_\GJ(\Omega R/c)$ (Arons
1981, 1983), the field $E{_\parallel}$ is screened at
distances beyond $s$. This condition together with (5.12) gives
a lower limit on the height, $H=H_i^{\rm min}$, for photoionization of
bound pairs near the pulsar surface to be important:
$$H_i^{\rm min}\simeq
(1.3\times10^2{\rm\,m})
\left({P\over0.1\,{\rm s}}\right)^{1/14}
\left({T_\S\over 10^6\,{\rm K}}\right)^{-3/7}\,.
\eqno(5.13)$$

\noindent In our estimate of the density of free pairs, $n_\pm$, cf.\
(5.12), the energy $2\Gamma_{\rm min}mc^2$ is adopted for both the
curvature photons and the created pairs. However, this energy is only the
threshold for pair creation, and the actual mean energy may be a factor
two or so higher. Alternative assumptions that give an estimate of the
maximum value, $H=H_i^{\rm max}$, of $H_i$ are that all curvature photons
are generated by primary particles at $s=0$, and all are emitted at
threshold. If we assume that the mean energy of each pair is
$4\Gamma_{\rm min}mc^2$, from (5.8) and (5.11) we have
$s\simeq H_i^{\rm max}\simeq3H_i^{\rm min}$.
We assume $H_i=2H_i^{\rm min}$.

Our estimate of the thickness of the polar gaps is (for $H_b>H_f$)
$$H=\,{\rm min}\,
\{H_b,\,{\rm max}\,[H_f,H_i]\}\,.
\eqno(5.14)$$

In summary, for free electron ejection from the polar cap, any increase in
$L^{\rm max}_{\rm nth}$ is rather small for all known pulsars, and it
cannot be account for strong nonthermal X-rays and $\gamma$-rays observed
from a few pulsars. The main reason for this is that $E_\parallel$
in the outflowing plasma is rather weak, and the thickness, $H_f$, of the
polar gaps is large even before taking into account its increase due
to the creation of bound pairs. Hence, the potential
$\Delta\varphi\sim E_\parallel H_f$ is modest, and with at best a modest
increase in $H$, it is not possible to increase $\Delta\varphi$
substantially.

\subsection{5.2. Limited ejection of electrons}

Now consider the opposite case where electrons do not flow freely from
the stellar surface. This requires that two conditions be satisfied
simultaneously: thermionic emission of electrons must be ineffective, and
ejection of electrons due to field emission must be ineffective. The
former of these conditions requires $T_\S<T_e$, cf.\ (2.13), and the
latter requires $E_{\parallel}<E_e$, cf.\ (2.14). Provided these
conditions are satisfied, $E_{\parallel}$ near the pulsar surface is of
order the vacuum value $E_{\parallel}^\RS\simeq(\Omega R/c)^{3/2}cB_\S$,
which is about $8\sqrt{3}(c/\Omega R)\sim 10^2$ -- $10^4$ times the
maximum value of $E_{\parallel}$ in the Arons model, cf.\ (3.11).

Assuming that $B$ is strong enough for the formation of bound pairs only
to occur, a self-consistent solution for the polar gap may be constructed
if field ionization of bound pairs is possible partially within the
polar gap (see below). This requires
$E^{\rm ion}_{\parallel}<E_{\parallel}^\RS$ within the polar gap. The
conditions $E^{\rm ion}_{\parallel}<E_{\parallel}^\RS$ and
$E_{\parallel}^\RS<E_e$ can be satisfied only for
$E^{\rm ion}_{\parallel}<E_e$. For $B\simeq(0.1$ -- $1)B_{\rm cr}$,
comparison of (2.14) and (4.29) with (4.33) shows
$E^{\rm ion}_{\parallel}\sim 0.1\,E_e$, implying that there can be a
range,  $E^{\rm ion}_{\parallel}<E_\parallel^\RS<E_e$, where escape from
the surface is ineffective and where field ionization of bound pairs is
possible.

When there is neither free ejection from the stellar surface
nor ionization of bound pairs, vacuum conditions apply. Then the
field $E_{\parallel}$ near the polar caps should be (Ruderman \and
Sutherland 1975):
$$E_{\parallel}(s,\rho)\simeq
{(\Delta R_p)^2-\rho^2
\over(\Delta R_p)^2}
\exp
\left(-{2s\over\Delta R_p}
\right)E_{\parallel}^\RS\,,
\eqno(5.15)$$

\noindent where $\rho$ is the distance to the magnetic axis.
Granted that $E_{\parallel}$ is unscreened, and so is given by (3.12), the
condition $E_{\parallel}^\RS=E_{\parallel}^{\rm ion}$ defines a minimum
period $P=P_1$ longer than which field ionization is unimportant:
$$P_1={2\pi R\over c}
\left({cB_\S\over E_{\parallel}^{\rm ion}}\right)^{2/3}
\simeq(0.5{\rm\,s})
\left({R\over 10^4\rm\,m}\right)
\left({B_\S\over 0.1\,B_{\rm cr}}\right)^{2/3}\,.
\eqno(5.16)$$

\noindent In the following discussion we assume $P<P_1$, so that field
ionization is important.

There is an inconsistency if one assumes that $E_\parallel$ is either
much smaller or much larger than $E_\parallel^{\rm ion}$. On the one hand,
suppose that ejection of electrons from the stellar surface is severely
limited, implying $E_\parallel\simeq E_\parallel^\RS\gg
E_\parallel^{\rm ion}$. For $P<P_1$ bound pairs created in the region
$0\le s<{1\over2}\Delta R_p$ are ionized before they escape from
this region. The resulting free positrons are accelerated toward the
surface of the star, producing an intense flux that heats the polar cap.
In the absence of ejection of electrons from the pulsar surface the
temperature at the polar cap would be (Ruderman \and Sutherland 1975)
$$T_\RS\simeq
(10^7{\rm\,K})
\left({B_\S\over 0.1\,B_{\rm cr}}\right)^{1/14}
\left({P\over P_1}\right)^{-3/14}\,.
\eqno(5.17)$$

\noindent It follows from (2.13) and (5.17) that, for both $B_\S=(0.1$ --
$1)B_{\rm cr}$ and $P<P_1$, $T_\RS$ is about an order of magnitude more
than $T_e$, as given by (2.13). This implies that thermionic emission (due
to the heating by positron bombardment) would produce an intense flux of
electrons from the polar cap for temperatures well below $T_\RS$. This
invalidates the original assumption that electron flow from the
surface is severely limited. On the other hand, if one assumes
$E_\parallel\ll E_{\parallel}^{\rm ion}$ for $P<P_1$, then field
ionization of bound pairs is negligible, and both the heating of the polar
caps by reversed particles and the screening of the field $E_{\parallel}$
are ineffective. As a result, one should have the vacuum field
$E_{\parallel}=E_{\parallel}^\RS$, which invalidates the assumption
$E_\parallel\ll E_{\parallel}^{\rm ion}$.

This inconsistency is avoided only if there is partial but not total
screening of $E_{\parallel}$ inside the polar gap. Specifically, this
field needs to be restricted to $E_{\parallel}\sim(1$ -- $2)E^{\rm
ion}_{\parallel}$. This is possible through a feedback mechanism. To see
that there is a feedback, suppose $E_{\parallel}$ decreases. This causes
the rate of ionization of the bound pairs to decrease, so that the
number of returning positrons heating the polar cap decreases, the
temperature of the polar cap decreases and the rate of thermionic emission
decreases. This leaves fewer (primary) electrons to screen
$E_{\parallel}$, which increases toward $E_{\parallel}^\RS$, providing the
feedback. An analogous argument leads to the conclusion that a postulated
increase in $E_{\parallel}$ implies a sequence of processes causing
$E_{\parallel}$ to decrease.

There is a self-consistent solution for the polar gap
with a strong magnetic field, $B(s=\Delta R_p)>0.1\,B_{\rm cr}$ or
$H_b>\Delta R_p$ (cf.\ Cheng \and Ruderman 1980). The surface is heated to
$T_\S\sim T_e$, required for marginally effective thermionic emission.
The number density of primary electrons, $n_{\rm prim}$, which is equal to
$n_\GJ$ in the limit $E^\RS_{\parallel}\gg E_\parallel^{\rm ion}$ of
complete ionization, is such that the difference $n_\GJ-n_{\rm prim}$ is
proportional to $E^\RS_{\parallel}-E_\parallel^{\rm ion}$. The potential
difference is just $E^{\rm ion}_{\parallel}$ times the height of the polar
gap, which is identified as the Ruderman-Sutherland value
${1\over2}\Delta R_p$. Thus this self-consistent model has
$$\displaylines{
\hfill
T_\S\simeq T_e\,,
\qquad
n_{\rm prim}\simeq n_\GJ
\left(1-{E^{\rm ion}_{\parallel}\over
E^\RS_{\parallel}}\right)\,,
\qquad
\Delta\varphi\simeq
{\textstyle{1\over 2}}
E^{\rm ion}_{\parallel}\Delta R_p\,,
\hfill\cr\hfill
E_{\parallel}\simeq \Theta [s]
\Theta [{\textstyle{1\over 2}}\Delta R_p-s]
E^{\rm ion}_{\parallel}\,,
\hfill
\llap{(5.18)}
\cr}$$

\noindent where $\Theta[x]$ is the step function, which is equal to unity
for $x>0$ and to zero for $x<0$.

To obtain (5.18) we assume that none of the parameters of the polar gap
depend on time. Alternatively it may be possible to find a time-dependent
gap solution, analogous to the sparking model of Ruderman and Sutherland
(1975). We have not explored this possibility in detail. Our
preliminarily considerations suggest that such a time-dependent model may
imply a somewhat higher (than our time-independent model) total power
carried away by both relativistic particles and radiation from the polar
gap into the pulsar magnetosphere.

Although photoionization of bound pairs occurs for $P<P_1$ it has little
effect on the model (5.18) provided the number density of ionized pairs,
$n_\pm$, at $s\simeq{1\over 2}\Delta R_p$ is smaller than
$n_\GJ-n_{\rm prim}\simeq
n_\GJ(E^{\rm ion}_{\parallel}/E^\RS_{\parallel})$. From (5.12) and
(5.18), the equality $n_\pm=n_\GJ-n_{\rm prim}$ defines a lower limit,
$P=P_2$, to the period:
$$P_2=(0.07{\rm\,s})
\left({T_\S\over 10^6\,{\rm K}}\right)^{4/11}
\left({B_\S\over 0.1\,B_{\rm cr}}\right)^{2/11}\,.
\eqno(5.19)$$

\noindent For $P<P_2$ the number density of free pairs is sufficient to
cause an additional screening, reducing $E_\parallel$ to $\ll
E_\parallel^{\rm ion}$. Thus the model (5.18) applies only for
$P_2<P<P_1$.

For $P_2<P<P_1$, (5.3) and (5.18) imply that the total power
carried away by both relativistic particles and radiation from
the polar gap into the pulsar magnetosphere is
$$L_p\simeq
\pi(\Delta R_p)^2n_{\rm prim}ce\Delta\varphi\simeq
{3\over2}{E^{\rm ion}_{\parallel}\over E^\RS_{\parallel}}
\left(1-{E^{\rm ion}_{\parallel}\over E^\RS_{\parallel}}
\right)\dot E_{\rm rot}\,,
\eqno(5.20)$$

\noindent where (e.g., Ostriker \and Gunn 1969)
$$\dot E_{\rm rot}\simeq
{2\pi\Omega^4R^6B^2_\S\over3\mu_0c^3}\simeq
(1.8\times10^{29}{\rm\,W})
\left({P\over0.1\,{\rm s}}\right)^{-4}
\left({B_\S\over0.1\,B_{\rm cr}}\right)^2
\eqno(5.21)$$

\noindent is the rate of rotation energy loss of the neutron star. The
maximum conceivable value of the energy flux in nonthermal particles is
$\dot E_{\rm rot}$. This maximum is achieved (to within a factor of order
unity) in the Sturrock model with
$L_p\simeq\pi(\Delta R_p)^2n_{\rm prim}ce\Delta\varphi$ in (5.20)
interpreted in terms of $n_{\rm prim}=n_\GJ$, cf.\ (2.1), and
$\Delta\varphi=\Delta\varphi_{\rm max}$, cf.\ (1.1). However, as already
remarked, the Sturrock model is not internally consistent. We now argue
that the modified model (5.18) can produce $L_p$ close to this maximum.

It is convenient to define the ratio $\eta^b_\gamma=L_p/\dot E_{\rm rot}$
of the spin-down power going into primary particles. In our modified model
this fraction is given by
$$\eta ^b_\gamma
={L_p\over \dot E_{\rm rot}}\simeq
{3\over 2}{E^{\rm ion}_{\parallel}\over
E^\RS_{\parallel}}
\left(1- {E^{\rm ion}_{\parallel}\over E^\RS_{\parallel}}
\right)
\quad{\rm at} \quad
P_2<P<P_1\,.
\eqno(5.22)$$

\noindent Equations (5.16) and (5.22) yield
$$\eta^b_\gamma\simeq
{3\over 2}\left({P\over P_1}\right)^{3/2}
\left[1-\left({P\over P_1}\right)^{3/2}\right]
\quad{\rm at}\quad
P_2<P<P_1\,.
\eqno(5.23)$$

\noindent The value of $\eta^b_\gamma$ has a maximum,
$\eta^b_\gamma={3/8}$, at $P=2^{-2/3}P_1\simeq0.6\,P_1$.

For comparison with (5.22) or (5.23), suppose that both ejection of
particles from the pulsar surface is limited and pairs are created free.
Then the fraction of the spin-down power radiated by pulsars is
(Ruderman \and Sutherland 1975; Cheng \and Ruderman 1980)
$$\eta_\gamma^f\simeq 1.5\times10^{-3}
\left({B_\S\over 0.1\,B_{\rm cr}}
\right)^{-8/7}
\left({P\over 0.1\,{\rm s}}\right)^{15/7}\,.
\eqno(5.24)$$

\noindent From (5.22) and (5.24) we can see that for $B_\S\simeq(0.1$ --
$1)B_{\rm cr}$ the ratio $\eta^b_\gamma/\eta^f_\gamma$ is $\sim 20$ at
$P=0.6\,P_1$ and $\sim10^2$ at $P=P_2$. Hence, at $P_2<P<P_1$ the
nonthermal luminosity of pulsars with strong magnetic fields is
considerably enhanced as a result of the creation of bound pairs instead
of free pairs. The nonthermal luminosity of such a pulsar with
$P\simeq0.6\,P_1$ may be comparable with the spin-down power, as in the
model of Sturrock (1971).

The model (5.18) breaks down for $P<P_2$ and for $P>P_1$. Consider the
lower limit, $P=P_2$. According to (5.12) with (2.1), the density of free
pairs resulting from field ionization increases with decreasing pulsar
period, $n_\pm\propto P^{-5}$, and for $P<P_2$ the number
density of ionized pairs dominates in the screening of $E_\parallel$,
thereby reducing $\Delta\varphi$ in (5.18) and hence $L_p$ according to
(5.19). The fraction of the spin-down power radiated by pulsars drops
sharply from $\sim\eta^b_\gamma$ for $P>P_2$ to $\sim\eta^f_\gamma$ for
$P\ll P_2$. For $P>P_1$ and $T_\S<T_e$, there is neither particle ejection
from the pulsar surface nor a cascade of pair production in the polar
gaps. In this case the nonthermal luminosity of pulsars is negligible.
Hence, for pulsars with strong magnetic fields, the death line should be
$P=P_1$ (cf.\ Chen \and Ruderman 1993). However, this predicted death
line is not supported by observation, as discussed further in section~5.5
below.

\subsection{5.3. ${\bf\Omega\cdot B}<0$  at the polar caps}

\par Ions tend to be extracted from the stellar surface in the polar caps
of pulsars by any parallel electric field in the case ${\bf\Omega\cdot
B}<0$. As for models with ${\bf\Omega\cdot B}>0$, it is convenient to
distinguish between the case where particles can flow freely from the
surface due to thermionic emission, and the case where flow from the
surface is limited. The discussion of these two cases closely parallels
those in sections~5.1 and~5.2, respectively.

\par Ions flow freely from the pulsar surface with density $\sim n_\GJ$
if the temperature of the polar cap satisfies $T_\S>T_i$, cf.\ (2.17). The
potential, $\Delta\varphi$, near the pulsar surface then cannot be much
greater than $\Delta\varphi_\RS$. The point is that for both
$\Delta\varphi>\Delta\varphi_\RS$ and $H>H_i$, cf.\ (5.13), the polar gap
is unstable, in the sense that a cascade of pair production develops. The
cascade is stopped only when either the potential becomes of order
$\Delta\varphi_\RS$ or the thickness of the polar gap becomes $\sim H_i$
when the photoionization of bound pairs is small. It follows from the
discussion in section 5.1 that for $H\sim H_i$, the potential across the
polar gap is not much greater than $\Delta\varphi_\RS$ for any known
pulsar. Any increase (due to the reduced screening resulting from the
pairs being bound rather than free) in the energy flux in particles is
at most modest. Hence, for free ejection of ions from the polar cap, the
creation of bound pairs in a strong magnetic field cannot be responsible
for the high-frequency radiation of pulsars.

Provided that the temperature of the polar cap, $T_\S$, without heating
by reversed electrons is smaller than $T_i$, thermionic emission of ions
is negligible. The $E{_\parallel}$ field distribution near the
polar cap is then given by (5.15). As for the case described
in section 5.2, for $P<P_1$, one has
$E{_\parallel}>E_\parallel^{\rm ion}$ and the potential $\Delta\varphi$
across the polar gap is substantially greater than the Ruderman-Sutherland
limit, $\Delta\varphi_\RS\simeq$ a few $\times10^{12}\rm\,V$. Such a gap is
unstable to pair production inside it. Indeed, if a pair is created inside
the
gap, the field $E{_\parallel}$ accelerates the positron out of the gap and
accelerates the electron toward the stellar surface. Since
$\Delta\varphi>\Delta\varphi_\RS$, one (or both) of these particles is
accelerated in the gap until it generates the curvature radiation which
is absorbed inside the gap to create secondary pairs. The total number of
such secondary pairs per primary particle is much greater than unity. With
$E{_\parallel}>E_\parallel^{\rm ion}$ these pairs are necessarily free. In
turn, the secondary pairs are accelerated and generate curvature photons
which are absorbed to create tertiary pairs and so on. This cascade ceases
if the polar cap is heated by reversed electrons so
that the surface temperature increases to $\sim T_i$. (One has
$T_i<T_\RS$, cf.\ (5.17), and the heating can increase the surface
temperature to $T_\RS$.) Then outflowing ions screen $E{_\parallel}$,
restricting it to $E{_\parallel}\sim(1$ -- $2)E_\parallel^{\rm ion}$.
Our modified polar gap, summarized in (5.18), then applies, with $T_e$
replaced by $T_i$ and with $n_{\rm prim}$ the sum of the density of
outflowing ions and of the positrons created in the polar gap. Thus,
(5.20) may be used to estimate the total power carried away by both
relativistic particles and radiation from the polar gap into the pulsar
magnetosphere.

\subsection{5.4. Interpretation of observational data}

Strong nonthermal radiation in X-rays and $\gamma$ rays has been observed
from a few radio pulsars (Gunji \et 1994; Ulmer 1994). Some
observational data and theoretical predictions on these pulsars are
presented in Table 1.

\par The strength of the magnetic field  given in Table~1 is estimated
from the slow down of the pulsar rotation ($B_\S\propto(P{\dot
P})^{1/2}$). This actually estimates the value of the
magnetic field near the light cylinder (the inner boundary of the
radiation zone). The strength of the magnetic field at the stellar surface
depends on the assumption about the structure of the magnetic field, which
is assumed dipolar. Thus, in Table 1, for each pulsar it is the dipole
component of the magnetic field, $B_\S$,  at the magnetic pole that
is given. These values of $B_\S$ are twice as strong as the estimates of
Michel (1991), whose $B$-values apply to the magnetic equator, rather
than to the magnetic poles which are more relevant for the polar gaps
(cf.\ Shapiro and Teukolsky 1983).

The values of $B_\S$ given in Table 1 are lower limits on the actual
surface fields. The estimated fields also depend on the angle, $i$,
between the dipole axis and the rotation axis, and the quoted values for
$i=\pi/2$ correspond to the minimum value of the dipole moment. More
importantly, the actual field may be much more complicated than a pure
dipole.  For example, suppose the magnetic field of a pulsar were purely
quadrupolar rather than purely dipolar; then to explain the same rate of
deceleration of the pulsar rotation, the surface magnetic field would
need to be $(c/\Omega R)\sim10^2$ -- $10^4$ times stronger. For
PSR1055--52 and Geminga, the dipolar components give fields smaller than
$0.1\,B_{\rm cr}$. If this were the actual field then pairs would be
created free.  In suggesting that bound pairs might cause an increase in
the pulsar luminosity, we need to assume a surface field stronger than
the dipolar value. The value $B_\S\sim6\times10^8\rm\,T$ is indicated in
parenthesis for these pulsars in Table 1.

\par From Table 1 we can see that the following inequalities hold
for all pulsars: $\eta^f_\gamma\ll\eta^b_\gamma$ and
$\eta_\gamma^{\rm obs}\la\eta^b_\gamma$. We conclude that the effect of
adiabatic conversion of the curvature $\gamma$-quanta into mutually bound
pairs in a strong magnetic field, $B>0.1\,B_{\rm cr}$, is able to increase
the nonthermal luminosities of pulsars sufficiently to account for the
observations.  However, the condition $P>P_2$, cf.\ (5.19), is not
satisfied for two of the pulsars listed, specifically for the Crab and the
Crab-like PSR 0540--69. Apart from these two, we conclude that
photoionization of bound pairs is unimportant in determining the polar-gap
structure, so that (5.23) applies. Thus the nonthermal high-frequency
radiation of these pulsars may be explained in terms of the modified
polar-gap model which takes account of the creation of bound pairs.

The Crab-like pulsars (PSR 0531+21 and PSR 0540--69) have periods
shorter than $P_2$. These are also young pulsars, which may not have had
time to cool sufficiently (to $T_\S<T_e$ or $T_\S<T_i$) for our model to
apply. The high-frequency radiation from these pulsars cannot be explained
in terms of our modified model. However, for these two Crab-like pulsars,
the outer-gap model of Cheng \et (1986a) seems satisfactory (e.g.,
Ulmer \et 1994).
Moreover, the $\gamma$-ray emission from the Crab-like pulsars may also
be explained in terms of the slot gap model of Arons (1983). However,
the slot gaps is an effective source of the energy for nonthermal
radiation only for dipole-like magnetic fields. Otherwise, the total
power carried away by both relativistic particles and radiation from the
slot gap into the pulsar magnetosphere is suppressed by a factor of
$\sim (R_c/R)(\Omega R/c)^{1/2}$ (Arons 1983). It is reasonable that
the radius of curvature of the magnetic field lines near the surface
is of order the stellar radius, $R_c\simeq R$, in which
case for the Crab pulsar the total power from the slot gap is even
smaller than the the total power from the polar gap in the standard
non-modified polar-gap models (Ruderman and Sutherland 1975; Arons 1979,
1981; Cheng and Ruderman 1980; Mestel {\it et al.} 1985). Therefore,
the outer-gap model of Cheng {\it et al.} (1986a) which is not so
sensitive to the structure of the magnetic field near the pulsar
is preferable as a model of the high-frequency radiation from
Crab-like pulsars.

\par In Table 1, to estimate the fraction, $\eta^b_\gamma$, of the
spin-down power, $\dot E_{\rm rot}$, radiated by pulsars in the
modified polar-gap model it is assumed that only one polar gap is a
source of both high-energy particles and radiation in the pulsar
magnetospheres. This assumption is reasonable for PSR 1706--44,
PSR 1509--58 and PSR 1055--52 which have only one $\gamma$-ray pulse
in their $\gamma$-ray light curves (e.g., Ulmer 1994). The
other pulsars have two pulses in their $\gamma$-ray light curves.
This indicates that in their magnetospheres both polar gaps operate as
particle accelerators. In this case the actual values of $\eta^b_\gamma$
are twice the values given in Table 1. This change strengthens the
conclusion that the modified polar-gap model is able to explain the
observed high-frequency luminosities of pulsars.

\par The distance to Geminga is particularly uncertain and the value
adopted ($150\rm\,pc$) is also indicated in parenthesis in Table~1. If
the distance is less than $\sim50\rm\,pc$, we have $\eta_\gamma^{\rm
obs}\la\eta_\gamma^f$, and the standard polar-gap models could describe
Geminga's $\gamma$ ray and X-ray emission adequately (Harding, Ozernoy
\and Usov 1993). The high-frequency luminosity of Geminga may be explained
in the modified polar-gap model if the distance to Geminga is up to
$\sim250\rm\,pc$.

Recently, Sturner and Dermer (1994) and Daugherty and Harding
(1994) have shown that double pulses with large interpulse separations
seen in the Crab, Vela, and Geminga can be produced by hollow cone
emission from a single pole if the rotation and magnetic axes are
nearly aligned. In these models the solid angle of $\gamma$-ray
emission is very small, and the standard polar gap models have no
problem supplying the power necessary to give the observed fluxes of
$\gamma$-rays. However, in this case the chance of observing any
given pulsar from the Earth is not more than $\sim 10^{-2}$. As
noted by Daugherty and Harding (1994), in spite of the poor statistics,
it seems hard to reconcile such a low probability of pulsar detection
with either the fraction of $\gamma$-ray pulsars observed among the
known supernova remnants, or the fraction which have a double-pulse
structure.

\subsection{5.5. The death line at $P=P_1$}

The prediction that $P=P_1$ should be the death line (no free pairs for
$P>P_1$) for pulsars with $B\ga0.1\,B_{\rm cr}$ is not supported by the
data. The distribution of radio pulsars essentially ignores our death
line. The radio emission is attributed to free pairs and hence it appears
that there is some source of free pairs even for pulsars with
$B>0.1\,B_{\rm cr}$ and $P>P_1$. There are several possibilities within
the framework of our model.
One suggestion concerns the decay of positronium from the excited state
formed from $\perp$-polarized photons (section~4.3). This decay should
produce some free pairs, due for example to two-photon decay (cf.\ Melrose
\and Kirk 1986) to the densely-packed states ($n_c\to\infty$) around zero
binding energy for the ground state ($n=n'=0$). Even a small branching
ratio for generation of free pairs may suffice to account for the apparent
lack of suppression of the radio emission. Two other possibilities for the
formation of some free pairs are suggested by a critical examination of
the assumption that the curvature photons emitted by the primary particles
cannot decay directly into free pairs (section~3.2). In our discussion in
section~3.2 we assume that {\it all\/} curvature photons are confined to a
forward cone with half angle $\sim1/\Gamma$. However, a small fraction,
$\sim1/\Gamma$, of the power is emitted outside this cone. Some of these
photons should satisfy the condition for decay into free pairs at the
point of emission, and these photons should be a source of some free
pairs. Another possibility follows from the fact that for any radiation
process that produces photons with energy above the threshold for pair
creation, the analogous process in which the photon is replaced by a pair
is allowed. Thus direct curvature emission of pairs is another possible
source of free pairs. Photon-photon production of pairs is yet a further
possibility. The efficacy of these processes needs to be explored, but
is subject to the major uncertainty that the number of free pairs
required to account for the radio emission is poorly constrained.

We conclude that $P=P_1$ should be regarded as a death line for
high-frequency emission, but not necessarily for radio emission, for
which the luminosity is many orders smaller than the spin-down power.
However, our arguments on this point are speculative, and a quantitative
analysis is desirable.

\section{6. CONCLUSIONS AND DISCUSSION}

\par In this paper we consider pulsars with strong magnetic fields,
$B_\S\ga0.1\,B_{\rm cr}$, at their surfaces such that bound rather than
free pairs should form due to decay of $\gamma$ rays in polar gaps.
In discussing various relevant aspects of the underlying physical
processes, we make the following points:
\smallskip

\item{1).} The surface structure of magnetic neutron stars determines
whether particles flow freely from the surface, leading to a model of the
type discussed by Arons (1981, 1983),  or whether particles are strongly
bound to the surface, leading to a model of the type discussed by
Ruderman \and Sutherland (1975) and Cheng \and Ruderman (1980).
For pulsars with ${\bf\Omega\cdot B}>0$, the
requirement on the surface temperature, $T_\S$, for electrons to remain
strongly bound is $T_\S<T_e$, with $T_e$ given by (2.13). For pulsars
with ${\bf\Omega\cdot B}<0$, the requirement for ions to remain
strongly bound is $T_\S<T_i$, with $T_i$ given by (2.7).

\item{2).} In polar-gap models the primary particles emit curvature
photons which are initially at too small an angle to ${\bf B}$ to decay
into pairs, even when the curvature drift and the polarization drift are
included.

\item{3).} For such photons emitted near the stellar surface, $s=0$, the
curvature of dipolar field lines allows decay into pairs with a minimum
Lorentz factor $\Gamma_{\rm min}$ that decreases $\propto1/s$ with
increasing height, cf.\ (5.8).

\item{4).} The photons can be either $\parallel$-polarized or
$\perp$-polarized. If the magnetic field is strong enough,
$B\ga8\times10^8\rm\,T$, photon splitting tends to cause the
$\perp$-polarized photons to decay into $\parallel$-polarized photons.

\item{5).} In a moderately strong magnetic field,
$B\ga4\times10^8\rm\,T$, photons evolve into a bound pair
(positronium) rather than decaying into a free pair.
The $\parallel$-polarized produce positronium in its ground state, and
the $\perp$-polarized produce positronium in an excited state that quickly
decays to the ground state through gyromagnetic emission (actually a
spin-flip transition).

\item{6).} The positronium atoms can be destroyed (a)~by field ionization
for $E_\parallel\ga E_\parallel^{\rm ion}$, with $E_\parallel^{\rm ion}$
given by (4.29), or (b)~by photoionization due to thermal radiation from
the neutron star, with mean free path given by (4.32).

\smallskip
\noindent
In applying these properties to polar-gap models for pulsars we
distinguish between two types of model mentioned in point 1).\ above:
models in which particles flow freely from the surface, implying
$E_\parallel=0$ at the surface (e.g., Arons 1981, 1983), and models in
which particles are tightly bound to the surface, so that $E_\parallel$
has the vacuum value just above the surface (e.g., Ruderman \and
Sutherland 1975). The parallel electric field in the former,
$E_\parallel^\A$ given by (3.11), is much smaller that that in the latter,
$E_\parallel^\RS$ given by (3.12), and there is a corresponding difference
in the potential drop, $\Delta\varphi\sim E_\parallel H$, across the polar
gap of height $H$. Our particular interest is in the energy flux in
primary particles, which is assumed to determine the nonthermal,
high-frequency luminosity (e.g., Harding \and Daugherty 1993). The energy
flux in primary particles is $n_{\rm prim}c\pi\Delta R_p^2e\Delta\varphi$,
where $n_{\rm prim}$ is the number density of primaries and $\Delta
R_p\simeq(\Omega R/c)^{1/2}R$ is the radius of the polar cap. For $n_{\rm
prim}\sim n_\GJ$, cf.\ (2.1), and $\Delta\varphi\sim\Delta\varphi_{\rm
max} ={1\over 2}E^\RS_\parallel\Delta R_p$,
cf.\ (1.1), this is of order the spin-down power of the pulsar.
However, Arons-type models have a small $E_\parallel^\A\ll
E_\parallel^\RS$, Ruderman-Sutherland-type models have small $H\ll
\Delta R_p$, and both have $\Delta\varphi \ll\Delta\varphi_{\rm max}
$, so that the power in primary particles is severely limited.
The formation of positronium, rather than free pairs tends to reduce
the screening of $E_\parallel$, and hence to increase $H$.

We find the following specific results for models in which bound pairs
are formed.

\smallskip

\item{7).} There is at most a modest increase in the power in primary
particles in Arons-type models, and even after modification to include
positronium formation, these models cannot account for the observed
high-frequency emission.

\item{8).} In the Ruderman-Sutherland-type models, the formation of bound
pairs can have a large effect on the model, greatly increasing the height
of the polar gap. This model requires $T_\S<T_e$ or $T_\S<T_i$, cf.\
point 1).\ above, and then photoionization of the bound pairs is
unimportant.

\item{9).} For $P>P_1$, cf.\ (5.16), corresponding to
$E_\parallel^\RS<E_\parallel^{\rm ion}$, cf.\ point 6).\ above, there is
no field ionization of the bound pairs. This suggests that $P=P_1$ should
be the death line for pulsars with $B>0.1\,B_{\rm cr}$.
Observationally, radio pulsars seem to ignore this death line. In
section~5.5 we suggest processes that should produce some free
pairs even for $P>P_1$, and note that only a modest number of pairs is
required to account for the radio emission.

\item{10).} For $P_2<P<P_1$ we argue for a self-consistent,
time-independent model in which field ionization generates just enough
free pairs for the reverse flux of primary particles to heat the polar cap
sufficiently to maintain thermionic emission at the rate required for
partial screening to reduce $E_\parallel$ from $E_\parallel^\RS$ to $\sim
E_\parallel^{\rm ion}$. This model is summarized in (5.18). The modified
model is only weakly dependent on whether electrons or ions are pulled
from the stellar surface; for ions, replace $T_e$ in (5.18) by $T_i$.

\item{11).} For $P<P_2$, cf.\ (5.19), the number density of pairs from the
field ionization exceeds the number density of primaries, and with
decreasing $P$ the model rapidly approaches that without bound-pair
formation.

\item{12).} In the modified model the ratio, $\eta^b_\gamma$, of the power
in primary particles to the spin-down power, cf.\ (5.23), can be close to
unity. In this sense the model is similar to that of Sturrock (1971), but
without the internal inconsistency in the Sturrock model.
\smallskip
\noindent
In summary, the most important modifications due to the formation of
bound pairs rather than free pairs is for pulsars with strong surface
magnetic fields, $B_\S\ga4\times10^8\rm\,T$, with cool surfaces,
$T_\S\la5\times10^5\rm\,K$, and periods in the range $P_2<P<P_1$. The
model proposed here could then account for the efficient high-frequency
emission.

The high-frequency pulsars are (Table~1) PSR 0531+21 (the Crab), PSR
0540--69, PSR 0833--45 (Vela), PSR 1706--44, PSR 1509--58, PSR 1055--52 and
Geminga. Of these, the middle three satisfy all the requirements for our
modified model, but the first two (Crab-like) pulsars have periods $P<P_2$
and the last two have surface magnetic fields $B_\S<4\times10^8\rm\,T$.
For the Crab-like pulsars, the photoionization of bound pairs inside the
polar gaps should be strong, and the modified polar-gap model is then not
applicable. The outer-gap model of (Cheng, Ho \and Ruderman 1986a) seems a
viable alternative for these Crab-like pulsars (e.g., Ulmer \et 1994).
We argue that the magnetic field is likely to have nondipolar components
and these are not included in the estimate of $B_\S$ in Table~1. With
plausible values for a nondipolar component, the model may also apply to
PSR 1055--52 and Geminga. Recently, Usov (1994) has shown that the
outer-gap model of Cheng, Ho \and Ruderman (1986b) is inconsistent with the
available data on Geminga, implying a polar-gap model. For Geminga the
actual distance is important in estimating whether or not the modified
polar-gap model is viable, as discussed in section~5.4. Alternatively,
Harding, Ozernoy \and Usov (1993) argued that, provided the distance to
Geminga is not more than $\sim40$ -- $50\rm\,pc$, it is possible to
explain both the X-rays and $\gamma$ rays in terms of the polar gap model
of Arons (1979, 1981).

We discuss some other details of the application to high-frequency
pulsars elsewhere (Usov \and Melrose 1994). We conclude by commenting on
the polarization of the high-frequency radiation. According to 4).\ above,
for $B_\S\ga0.2\,B_{\rm cr}$, most of the $\perp$-polarized photons with
$\varepsilon_\gamma\la10^2\rm\,MeV$ produced by curvature mechanism near
the pulsar surface, are split and transformed into $\parallel$-polarized
photons before the pair creation threshold is reached (section 4.2). As a
result, the $\gamma$-ray emission recorded from the pulsar at energies
$\varepsilon_\gamma\la10^2\rm\,MeV$ should be linearly polarized. Near the
maximum of the light curve, the $\gamma$-ray polarization may be as high
as 100 per cent. The degree of $\gamma$-ray polarization should decrease
toward the edges of the $\gamma$-ray pulses. This is because the emission
zone recedes from the neutron star surface as the phase of $\gamma$-ray
emission shifts from the maximum of the light curve, implying that the $B$
value in the line of sight to the region of $\gamma$-ray generation
decreases. By observing the polarization of the $\gamma$-ray emission of
pulsars it would be possible to estimate the strength of the magnetic
field near the pulsar surface.

\vfill\eject

\section{APPENDIX A: PITCH-ANGLES IN THE DRIFT FRAME}

To estimate the angle $\tilde\psi$ between the particle
velocity $\tilde{\bf v}$ and the magnetic field $\tilde{\bf B}$
for a relativistic electrons in its lowest Landau level,
let us first consider the nonrelativistic motion of electrons
in crossed ${\bf B}$ and ${\bf E}_\perp(t)$  fields. The polarization
drift (e.g., Schmidt 1966; Landau \and Lifshitz 1971) implies a drift
$(e/m\omega_\B^2)\,d{\bf E}_\perp(t)/dt$. For a relativistic electron in
its lowest Landau orbital, the polarization drift may be obtained by
making a Lorentz transformation from the rest frame. The momentum in
the drift frame is
$$\tilde p_{\rm pd}=
{e\over\omega_\B^2}
\left({dE_\perp\over dt}\right)_c\,,
\eqno({\rm A}1)$$

\noindent where $(dE_\perp/dt)_c$ is to be evaluated in the frame in which
the electron is at rest on the average. With $(E_\perp)_c=\Gamma E_\perp$
and $(dz)_c=\Gamma^{-1}dz$ (e.g., Landau \and Lifshitz 1971), we have
$$\left({dE_\perp\over dt}\right)_c=
c\Gamma^2{dE_\perp\over dz}\,.
\eqno({\rm A}2)$$

\par From (A1) and (A2), we obtain
$$\tilde\psi\simeq
{\tilde p_{\rm pd}\over\tilde p}\le
{m\Gamma \over eB^2}{dE_\perp\over dz}\,,
\eqno({\rm A}3)$$

\noindent where $\tilde p\simeq mc\Gamma$ is the particle momentum in the
drift frame.

\vfill\eject

\section{APPENDIX B: MINIMUM ENERGY OF CREATED PAIRS}

Let us consider a neutron star with a centered magnetic dipole moment,
$\mu$. In cylindrical coordinates $z$, $\rho$, $\varphi$, the
components of ${\bf B}$ near the polar cap are
$$B_z\simeq{\mu_0\over4\pi}\,{2\mu\over z^3},
\qquad
B_\rho\simeq{\mu_0\over4\pi}\,{3\mu\rho\over z^4},
\qquad
B_\varphi=0\,,
\eqno({\rm B}1)$$

\noindent where $\mu$ is the magnetic moment, $z$-axis is directed along
the magnetic axis, $z$ is measured from the neutron star center, $\rho$ is
the distance to the magnetic axis, and $\varphi$ is the azimuthal
coordinate. The magnetic field lines are determined by integrating
$${dz\over B_z}={d\rho \over B_\rho}\,,
\eqno({\rm B}2)$$

\noindent which with (B1) and (B2) gives
$$\rho=\rho_0(z/z_0)^{3/2}\,,
\eqno({\rm B}3)$$

\noindent where $\rho_0/z_0^{3/2}$ is a constant. (The exact result is
$\rho=(\rho_0/z_0^{3/2})(\rho^2+z^2)^{3/2}$.)

\par For a curvature photon which is emitted
at the point with $z=z_0$ and $\rho=\rho_0$
along the magnetic field, ${\bf K}\parallel{\bf B}$,
the angle between the wave
vector ${\bf K}$ and the magnetic field ${\bf B}$ in the process of
the photon propagation is
$$\vartheta (z)=\left({d\rho \over dz}\right)_0
-\left({d\rho\over dz}\right)=
{3\over2}{\rho_0(z^{1/2}-z_0^{1/2})\over
z_0^{3/2}}\,.
\eqno({\rm B}4)$$

\noindent Taking into account that $\rho_0\le(\Omega R/c)^{1/2}R$
for open magnetic field lines near the polar caps of pulsars,
$z\geq R$ and $z-R\ll R$, (B4) gives
$$\vartheta(z)\le
{3\over4}
\left({\Omega R\over c}\right)^{1/2}
{z-R\over z}\,.
\eqno({\rm B}5)$$

\noindent At a distance $s$ from pulsar surface, with $s\simeq z-R$, we
have
$$\vartheta(s)\le
{3\over4}
\left({\Omega R\over c}\right)^{1/2}
{s\over R}\,.
\eqno({\rm B}6)$$

\par If the energy of curvature photons, $\varepsilon_\gamma$, is smaller
than
$$\varepsilon_{\rm min}=
2mc^2/\vartheta \,,
\eqno({\rm B}7)$$
\noindent the process of pair creation by these photons is kinematically
forbidden. From (B6) and (B7) it follows that the Lorentz
factor of particles which can be created at the distance $s$ exceeds
$$\Gamma_{\rm min}=
{\varepsilon_{\rm min}\over 2mc^2} = {4\over 3}
\left({c\over\Omega R}\right)^{1/2}
{R\over s}\,.
\eqno({\rm B}8)$$

\vfill\eject
\ctrline{\bf REFERENCES}

\def\aa {{\tf  Astr.\ Ap}.\ }

\def\apj{{\tf  Astrophys.\ J}.\ }
\def\apjs{{\tf  Astrophys.\ J.\ Suppl}.\ }

\def\ass{{\tf  Astrophys.\ Space Sci}.\ }

\def\ajp{{\tf  Aust.\ J.\ Phys}.\ }

\def\mnras{{\tf  Mon.\ Not.\ R.\ Astron.\ Soc}.\ }

\def\prev{{\tf  Phys.\ Rev}.\ }

\def\prl{{\tf  Phys.\ Rev.\ Lett}.\ }

\def\rmp{{\tf Rev.\ Mod.\ Phys}.\ }

\def\sovaj{{\tf Soviet Astron}.\ }

\def\usp{{\tf Soviet Phys.\ Usp}.\ }

\def\ssr{{\tf  Space Sci.\ Rev}.\ }

\def\cambridge{Cambridge University Press}

\def\reidel{D.~Reidel (Dordrecht)}

\def\sieber{in W.~Sieber and R.~Wiele\-binski (eds) {\tf Pulsars}.\
    \ IAU Symposium No.~95, \reidel}

\def\tf{\it}
\def\vf{\bf}

\def\ref#1#2#3#4{\hang#1, #2, #3, #4.}
\def\book#1#2#3#4{\hang#1, #2, {\tf#3}, #4.}

\ref
{Abrahams, A.M., \and Shapiro, S.L.}
{1991}
{Equation of state in a strong magnetic field:
finite temperature and gradient corrections}
{\apj {\vf 374}, 652--667}

\ref
{Adler, S.L.}
{1971}
{Photon splitting and photon dispersion in a strong magnetic field}
{{\tf Ann.\ Phys}.\ {\vf 67}, 599--647}

\ref
{Adler, S.L., Bahcall, J.N., Callan, C.G., \and Rosenbluth, M.N.}
{1970}
{Photon splitting in a strong magnetic field}
{\prl {\vf 25}, 1061--1065}

\ref
{Arons, J.}
{1979}
{Some problems of pulsar physics}
{\ssr {\vf 24}, 437--510}

\ref
{Arons, J.}
{1981}
{Pair creation above pulsar polar caps: steady flow in the
surface acceleration zone and polar cap X-ray emission}
{\apj {\vf 248}, 1099--1116}

\ref
{Arons, J.}
{1983}
{Pair creation above pulsar polar caps:
Geometrical structure and energetics of slot gaps}
{\apj {\vf 266}, 215--241}

\ref
{Arons, J., \and Scharlemann, E.T.}
{1979}
{Pair formation above polar caps:
structure of the low altitude acceleration zone}
{\apj {\vf 231}, 854--79}

\ref
{Baring, M.G.}
{1991}
{Signatures of magnetic photon splitting in gamma-ray burst spectra}
{\aa {\vf 249}, 581--588}

\ref
{Beskin, V.S.}
{1982}
{Dynamic screening of the acceleration region in the
magnetosphere of a pulsar}
{\sovaj {\vf 26}, 443--446}

\ref
{Beskin, V.S., Gurevich, A.V., \and Istomin, Ya.N.}
{1986}
{Physics of pulsar magnetospheres}
{\usp {\vf29}, 946--970}

\ref
{Bhatia, V.B., Chopra, N., \and Panchapakesan, N.}
{1988}
{The effect of photon capture and field ionisation
on high magnetic field pulsars}
{\ass {\vf 150}, 181--188}

\ref
{Bhatia, V.B., Chopra, N., \and Panchapakesan, N.}
{1992}
{Photodissociation in strong magnetic fields
and application to pulsars}
{\apj {\vf 388}, 131--137}

\ref
{Bialynicka-Birula, Z., \and Bialynicki-Birula, I.}
{1970}
{Nonlinear effects in quantum electrodynamics.
Photon propagation and photon splitting
in an external field}
{\prev {\vf D2}, 2341--2345}

\ref
{Bogovalov, S.V., \and Kotov, Yu.D.}
{1989}
{Electromagnetic cascade in a pulsar magnetosphere, and
the processes on the neutron star surface}
{{\tf Soviet Astron.\ Letters\/} {\vf 15}, 185--189}

\ref
{Chen, K., \and Ruderman, M.A.}
{1993}
{Pulsar death lines and death valley}
{\apj {\vf 402}, 264--270}

\ref
{Chen, H.-H., Ruderman, M.A., \and Sutherland, P.G.}
{1974}
{Structure of solid iron in superstrong
neutron-star magnetic fields}
{\apj {\vf 191}, 473--477}

\ref
{Cheng, K.S., Ho, C., \and Ruderman, M.A.}
{1986a}
{Energetic radiation from rapidly spinning pulsars. I. Outer
magnetosphere gaps}
{\apj {\vf 300}, 500--521}

\ref
{Cheng, K.S., Ho, C., \and Ruderman, M.A.}
{1986b}
{Energetic radiation from rapidly spinning pulsars. II. Vela and
Crab}
{\apj {\vf 300}, 522--539}

\ref
{Cheng, A.F., \and Ruderman, M.A.}{1977}
{Pair-production discharges above pulsar polar caps}
{\apj {\vf 214}, 598--606}

\ref
{Cheng, A.F., \and Ruderman, M.A.}
{1980}
{Particle acceleration and radio emission above pulsar polar
caps}
{\apj {\vf 235}, 576--586}

\ref
{Daugherty, J.K., \and Harding, A.K.}
{1994}
{Polar cap models of gamma-ray pulsar: emission from single
poles of nearly aligned rotators}
{\apj {\vf 429}, 325--330}

\ref
{Daugherty, J.K., \and Ventura, J.}
{1978}
{Absorption of radiation by electrons
in intense magnetic fields}
{\prev {\vf D18}, 1053--1067}

\ref
{Dermer, C.D.}
{1990}
{Compton scattering in strong magnetic fields and the continuum
spectra of gamma-ray bursts: basic theory}
{\apj {\vf 360}, 197--214}

\ref
{Erber, T.}
{1966}
{High-energy electromagnetic conversion processes
in intense magnetic fields}
{\rmp {\vf 38}, 626--659}

\ref
{Fitzpatrick, R., \and Mestel, L.}
{1988a \& b}
{Pulsar electrodynamics -- I.\ \& II.}
{\mnras {\vf 232}, 277--302 \& 303--321}

\ref
{Flowers, E.G., Lee, J.-F., Ruderman, M.A., Sutherland, P.G.,
Hillebrandt, W., \and M\"uller, E.}
{1977}
{Variation calculation of ground-state energy of iron
atoms and condensed matter in strong magnetic fields}
{\apj {\vf 215}, 291--301}

\ref
{Fushiki, I., Gudmundsson, E.H., \and Pethick, C.J.}
{1989}
{Surface structure of neutron stars with high magnetic fields}
{\apj {\vf 342}, 958--975}

\ref
{Ginzburg, V.L., \and Usov, V.V.}
{1972}
{Concerning the atmosphere of magnetic neutron stars (pulsars)}
{JETP Letters, 15, 196--198}

\ref
{Goldreich, P., \and Julian, W.H.}
{1969}
{Pulsar electrodynamics}
{\apj {\vf 157}, 869--880}

\book
{Gopal, E.S.R.}
{1974}
{Statistical Mechanics and Properties of
Matter}{(New York: Wiley)}

\ref
{Gunji, S. {\it et al.}}
{1994}
{Observation of pulsed hard X-rays/$\gamma$-rays from PSR 1509-58}
{\apj {\vf 428}, 284--291}

\ref
{Halpern, J.P., \and Holt, S.S.}
{1992}
{Discovery of soft X-ray pulsations from the $\gamma$-ray source Geminga}
{{\tf Nature\/} {\vf 357}, 222--224}

\ref
{Harding, A., \and Daugherty, J.K.}
{1993}
{Pulsar gamma-ray emission in the polar cap cascade model}
{in Van Riper, K.A., Epstein, R., \and Ho, C.\ (eds)
{\tf Isolated Pulsars}
\cambridge, pp.\ 279--286}

\ref
{Harding, A., Ozernoy, L.M., \and Usov, V.V.}
{1993}
{Geminga, origins of its X-ray and gamma-ray emission}
{\mnras {\vf 265}, 921--925}

\ref
{Hayward, E.}
{1965}
{Photonuclear reactions}
{in MacDonald, N., (ed.) {\tf Nuclear Structure and Electromagnetic
Interaction}, (Edinburgh:  Oliver \and Boyd) pp.\ 141--209}

\ref
{Herold, H., Ruder, H., \and Wunner, G.}
{1982}
{Cyclotron emission in strongly magnetized plasmas}
{\aa {\vf 115}, 90--96}

\ref
{Herold, H., Ruder, H., \and Wunner, G.}
{1985}
{Can $\gamma$ quanta really be captured by pulsar magnetic
fields?}
{\prl {\vf 54}, 1452--1455}

\book
{Jackson, J.D.}
{1975}
{Classical Electrodynamics}
{(New York: Wiley)}

\ref
{Jones, P.B.}
{1978}
{Particle acceleration at the magnetic poles of a neutron star}
{\mnras {\vf 184}, 807--827}

\ref
{Jones, P.B.}
{1979}
{Pair production on the pulsar magnetosphere}
{\apj {\vf 228}, 536--540}

\ref
{Jones, P.B.}
{1985}
{Density-functional calculations of the cohesive energy
of condensed matter in very strong magnetic fields}
{\prl {\vf 55}, 1338--1340}

\ref
{Jones, P.B.}
{1986}
{Properties of condensed matter in very strong magnetic fields}
{\mnras {\vf 218}, 477--487}

\ref
{Klepikov, N.P.}
{1954}
{Radiation of photons and electron-positron pairs in a magnetic field}
{{\tf Zh.\ Eksp.\ Teor.\ Fiz}.\ {\vf 6}, 19--35}

\ref
{Koribalski, B., Johnston, S., Weisberg, W.M., \and Wilson, W.}
{1995}
{HI line measurements of eight southern pulsars}
{\apj (in press)}

\book
{Landau, L.D., \and Lifshitz, E.M.}
{1971}
{The Classical Theory of
Fields}{(Oxford: Pergamon)}

\ref
{Levinson, A., \and Eichler, D.}
{1993}
{Baryon purity in cosmological gamma-ray bursts
as a manifestation of event horizons}
{\apj {\vf 418}, 386--390}

\ref
{Lieb, E.H.}
{1981}
{Thomas-Fermi and related theories of atoms and molecules}
{\rmp {\vf 53}, 603--641}

\ref
{Loudon, R.}
{1959}
{One-dimensional hydrogen atom}
{{\tf Am.\ J.\ Phys}.\ {\vf 27}, 649--655}

\ref
{Melrose, D.B.}
{1983}
{Quantum
electrodynamics in strong magnetic fields.
II Photon fields and interactions}
{\ajp {\vf 36}, 775--798}

\ref
{Melrose, D.B., \and Kirk, J.G.}
{1986}
{Two-photon emission in X-ray pulsars 1. Basic formulas}
{\aa {\vf 156}, 268--276}

\ref
{Melrose, D.B., \and Parle, A.J.}
{1983}
{Quantum
electrodynamics in strong magnetic fields. I Electron
states}
{\ajp {\vf 36}, 755--774}

\ref
{Melrose, D.B., \and Usov, V.V.}
{1994}
{Bound pair creation in polar gaps and X-ray
 and gamma-ray emission from radio pulsars}
{\apj (submitted)}

\ref
{Melrose, D.B., \and Zheleznyakov, V.V.}
{1981}
{Quantum theory of cyclotron emission and the X-ray line in Her
X-1}
{\aa {\vf 95}, 86--93}

\ref
{Mestel, L.}
{1981}
{Structure of the pulsar magnetosphere}
{\sieber, pp.\ 9--23}

\ref
{Mestel, L.}
{1993}
{Pulsar magnetospheres}
{in Blandford, R.D., Hewish, A., \and Mestel, L., (eds)
{\tf Pulsars as Physics Laboratories}, Oxford University Press,
pp.\ 93--104}

\ref
{Mestel, L., Robertson, J.A., Wang, Y.-M., \and Westfold, K.C.}
{1985}
{The axisymmetric pulsar magnetosphere}
{\mnras {\vf 217}, 443--484}

\book
{M\'esz\'aros, P.}
{1992}
{High-Energy Radiation from Magnetized Neutron Stars}
{(Chicago: Univ.\ of Chicago Press)}

\ref
{Michel, F.C.}
{1975}
{Composition of the neutron star surface in pulsar models}
{\apj {\vf 198}, 683--685}

\book
{Michel, F.C.}
{1991}
{Theory of Neutron Star Magnetospheres}
{(Chicago: Univ.\ Chicago Press)}

\ref
{M\"uller, E.}
{1984}
{Variation calculation of iron and helium atoms and
molecular chains in superstrong magnetic fields}
{\aa {\vf 130}, 415--418}

\ref
{Neuhauser, D., Koonin, S.E., \and Langanke, K.}
{1987}
{Hartree-Fock calculations of atoms and molecular chains in
strong magnetic fields}
{{\tf Phys.\ Rev}.\ {\vf A36}, 4163--4175}

\ref
{Neuhauser, D., Langanke, K., \and Koonin, S.E.}
{1986}
{Hartree-Fock calculations of atoms and molecular chains in
strong magnetic fields}
{\prev {\vf A33}, 2084--2086}

\ref
{Ochelkov, Yu.P., \and Usov, V.V.}
{1980}
{Curvature radiation of relativistic particles
in the magnetosphere of pulsars}
{\ass {\vf 69}, 439--460}

\ref
{Ogata, S., \and Ichimaru, S.}
{1990}
{Electric and thermal conductivities of quenched neutron star crusts}
{\apj {\vf 361}, 511--513}

\ref
{Ostriker, J.P., \and Gunn, J.E.}
{1969}
{On the nature of pulsars: I. Theory}
{\apj {\vf 157}, 1395--1417}

\ref
{Ozernoy, L.M., \and Usov, V.V.}
{1977}
{The nature of pulsar gamma rays}
{\sovaj {\vf 21}, 425--431}

\ref
{Page, D., \and Applegate, J.H.}
{1992}
{The cooling of neutron stars by direct URCA process}
{\apj {\vf 394}, L17--L20}

\ref
{Pavlov, G.G., \and M\'esz\'aros, P.}
{1993}
{Finite-velocity effects on atoms in strong magnetic fields
and implications for neutron star atmospheres}
{\apj {\vf 416}, 752--761}

\ref
{Rosen, L.C., \and Cameron, A.G.W.}
{1972}
{Surface composition of magnetic neutron stars}
{\ass {\vf 15}, 137--152}

\ref
{Rozental, I.L., \and Usov, V.V.}
{1985}
{Cascade processes in the surface layers of pulsars}
{\ass {\vf 109}, 365--371}

\ref
{Ruderman, M.A.}
{1971}
{Matter in superstrong magnetic fields:
The surface of a neutron star}
{\prl {\vf 27}, 1306--1308}

\ref
{Ruderman, M.A., \and Sutherland, P.G.}
{1975}
{Theory of pulsars: polar gaps, sparks, and coherent
microwave radiation}
{\apj {\vf 196}, 51-72}

\ref
{Schiff, L.I., \and Snyder, H.}
{1939}
{Theory of the quadratic Zeeman effect}
{\prev {\vf 55}, 59--63}

\book
{Schmidt, G.}
{1966}
{Physics of High Temperature Plasmas}
{(New York:Academic Press)}

\ref
{Shabad, A.E.}
{1975}
{Photon dispersion in a strong magnetic field}
{{\tf Ann.\ Phys}.\ {\vf 90}, 166--195}

\book
{Shabad, A.E.}
{1992}
{Polarization of the Vacuum and a Quantum
Relativistic Gas in an External Field}
{(New York: Nova Science Publishers)}

\ref
{Shabad, A.E., \and Usov, V.V.}
{1982}
{$\gamma$-Quanta capture by magnetic field and
pair creation suppression in pulsars}
{{\tf Nature\/} {\vf 295}, 215--217}

\ref
{Shabad, A.E., \and Usov, V.V.}
{1984}
{Propagation of $\gamma$-radiation in strong magnetic
fields of pulsars}
{\ass {\vf 102}, 327--358}

\ref
{Shabad, A.E., \and Usov, V.V.}
{1985}
{Gamma-quanta conversion into positronium atoms
in a strong magnetic field}
{\ass {\vf 117}, 309--325}

\ref
{Shabad, A.E., \and Usov, V.V.}
{1986}
{Photon dispersion in a strong magnetic field with
positronium formation: theory}
{\ass {\vf128}, 377--409}

\book
{Shapiro, S.L., \and Teukolsky, S.A.}{1983}
{Black Holes, White Dwarfs
and Neutron Stars: The Physics of Compact Objects}
{(New York:
Wiley)}

\ref
{Shibata, S.}
{1991}
{Magnetosphere of the rotation-powered pulsar: a DC circuit
model}
{\apj {\vf 378}, 239--254}

\ref
{Skjervold, J.E., \and \"Ostgaard, E.}
{1984a}
{Hydrogen and helium atoms in superstrong magnetic fields}
{{\tf Phys.\ Scripta\/} {\vf 29}, 448--455}

\ref
{Skjervold, J.E., \and \"Ostgaard, E.}
{1984b}
{Heavy atoms in superstrong magnetic fields}
{{\tf Phys.\ Scripta\/} {\vf 29}, 543--550}

\ref
{Slattery, W.L., Doolen, G.D., \and DeWitt, H.E.}
{1980}
{Improved equation of state for the classical one-component plasma}
{\prev {\vf A21}, 2087--2095}

\ref
{Stoneham, R.J.}
{1979}
{Photon splitting in the magnetized vacuum}
{{\tf J. Phys}.\ {\vf A12}, 2187--2203}

\ref
{Sturner, S.L., \and Dermer, C.D.}
{1994}
{On the spectra and pulse profiles of gamma-ray pulsars}
{\apj {\vf 420}, L79--L82}

\ref
{Sturrock, P.A.}
{1971}
{A model of pulsars}
{\apj {\vf 164}, 529--556}

\ref
{Taylor, J.H., Manchester, R.N., \and Lyne, A.G.}
{1993}
{Catalog of 558 pulsars}
{\apjs {\vf 88}, 529--568}

\ref
{Toll, J.S.}
{1952}
{}
{dissertation, Princeton University}

\ref
{Tsai, W., \and Erber, T.}
{1974}
{Photon pair production in intense magnetic fields}
{{\tf Phys.\ Rev.\ D\/} {\vf 10}, 492--499}

\ref
{Ulmer, M.P.}
{1994}
{Gamma-ray observations of pulsars}
{\apjs {\vf 90}, 789--795}

\ref
{Ulmer, M.P., \et}
{1994}
{OSSE observations of the Crab pulsar}
{\apj {\vf 432}, 228--238}

\ref
{Usov, V.V.}
{1994}
{Radiation from Vela-like pulsars near the death line}
{\apj {\vf 427}, 394--399}

\ref
{Usov, V.V. \and Shabad, A.E.}
{1983}
{The decay of curvature gamma-ray photons
near a neutron star surface}
{{\tf Soviet Astron.\ Letters\/} {\vf 9}, 212--214}

\ref
{Usov, V.V., \and Shabad, A.E.}
{1985}
{Photopositronium in the magnetosphere of a pulsar}
{{\tf JETP Letters\/} {\vf 42}, 19--23}

\ref
{Van Riper, K.A.}
{1991}
{Neutron star thermal evolution}
{\apjs {\vf 75}, 449--462}

\ref
{Von Neumann, J., \and Wigner, E.}
{1964}
{On the behavior of eignenvalues in adiabatic processes}
{in R.S.\ Knox and A.\ Gold (eds) {\it Symmetry in Solid Sate},
(W.A. Benjamin, New York, pp.\ 167--172}

\ref
{Woosley, S.E., \and Baron, E.}
{1992}
{The collapse of white dwarfs to neutron stars}
{\apj {\vf 391}, 228--235}

\ref
{Wunner, G, \and Herold, H.}
{1979}
{Decay of positronium in strong magnetic fields}
{\ass {\vf 63}, 503--509}

\ref
{Zheleznyakov, V.V., \and Shaposhnikov, V.E.}
{1979}
{Absorption of curvature radiation}
{\ajp {\vf 32}: 49--59}

\vfill\eject

$$\vbox{
\tabskip=0pt
\halign{
\hfil#\hfil&
\hfil$\;#\;$\hfil&
\hfil$\;#\;$\hfil&
\hfil$\;#\;$\hfil&
\hfil$\;#\;$\hfil&
\hfil$\;#\;$\hfil&
\hfil$\;#\;$\hfil&
\hfil$\;#\;$\hfil&
\hfil$\;#\;$\hfil
\cr
\noalign{\hrule\vskip5pt}
Name   & P  &   B_\S   &   D   &      L_{X+\gamma}&
\dot E_{\rm rot}  &  \eta^{\rm obs}_\gamma&        \eta^f_\gamma &
\eta^b_\gamma \cr
\noalign{\vskip3pt}
&{\rm ms}&10^8{\rm\,T}&{\rm kpc}&10^{29}{\rm W}&
10^{29}{\rm W} &10^{-2}&10^{-2}&10^{-2} \cr
\noalign{\vskip5pt\hrule\vskip5pt}
PSR 0531+21 &   33  &  6.6   &   2    &    2.2&
450 &  0.5    &     0.01  &  1.7\cr
PSR 0540--69 &   50   &   9  &   55   & 0.9 &  150
&  0.6 &  0.02 & 2.3\cr
PSR 0833--45  &  89  &  6.8 &   0.5  &  0.084&   7
 & 1.2 & 0.08 & 4.6\cr
PSR 1706--44  & 102  &  6.3 &   1.5     & 0.084&  3.4
& 2.5 &  0.1 &    6\cr
&&& (2.8) & 0.3&& 8.7&&\cr
PSR 1509--58  & 150  &  31  &  4.2  &0.39&  20
& 2  & 0.04 & 3.6\cr
PSR 1055--52 &  197  &  2 & 0.9& 0.006& 0.03 &
20  &  1 & -\cr
&&&(1.8)& 0.024 & & 80 &&\cr
&& (6) & & & & & & 22\cr
Geminga   &    237  &  3.3 & (0.15)
& 0.003&0.035& 9  & 1&-\cr
&& (6) & & & & & & 27\cr
\noalign{\vskip5pt\hrule}
}
}$$

\noindent{Table~1}: {The properties of high-frequency pulsars with data
taken from Ulmer (1994). The distances to PSRs 1706--44, from
Taylor, Manchester \and Lyne (1993), and to 1055--52, from Koribalski \et
(1995) are indicated in parenthesis.}

\vfill\eject

\section{Figure Captions}

\bigskip

\noindent{Figure~1}: {The dispersion curves for photon, (4.5), and
positronium, (4.6), are shown in the absence of interaction (solid
curves). The interaction is taken into account schematically in the dotted
curves, for $\parallel$-polarized photons, and the dot-dashed curves,
are for $\perp$-polarized photons, near the relevant intersection points
(circles). The positronium states are defined by (4.7) with $P_x=\hbar
K_\perp$; the labels on the curves denote the quantum number $n_c$ and the
parity under reflection along the $z$-axis (superscripts). [After  Shabad
\and Usov 1986.]}

\bigskip

\noindent{Figure~2}: {As in Figure~1, with the solid lines denoting the
mixed ($\parallel$-polarization/positive-parity)
photon-positronium states. The dots denote higher order (in $n_c$)
states that merge into a continuum near $\varepsilon=mc^2$.
[After  Shabad \and Usov 1986.]}

\vfill\eject

\midinsert
\vskip5cm
\centerline{
\scaledillustration 188mm by 130mm (shabadusov1 scaled 600)}
\vskip5cm
\centerline{\bf Fig. 1}
%\vskip1pt
\endinsert

\vfill\eject

\midinsert
%\vskip1pt
\vskip5cm
\centerline{
\scaledillustration 193mm by 130mm (shabadusov2 scaled 600)}
\vskip5cm
\centerline{\bf Fig. 2}
%\vskip1pt
\endinsert

\bye